\documentclass[journal=jacsat,manuscript=article]{achemso}
\setkeys{acs}{etalmode=truncate,maxauthors=10}
\mciteErrorOnUnknownfalse

\usepackage[version=3]{mhchem} 
\usepackage{amssymb}
\usepackage{xcolor}
\usepackage[colorlinks=true,breaklinks=true,linkcolor=black,citecolor=green,urlcolor=magenta]{hyperref}
\usepackage{siunitx}
\usepackage{graphicx}

\usepackage{xcolor}
\usepackage[capitalise]{cleveref}
\usepackage{algorithmicx}
\usepackage{algorithm}
\usepackage[noend]{algpseudocode}
\usepackage[normalem]{ulem}
\usepackage{makecell}

\algnewcommand\algorithmicinput{\textbf{Input:}}
\algnewcommand\Input{\item[\algorithmicinput]}

\algnewcommand\algorithmicoutput{\textbf{Output:}}
\algnewcommand\Output{\item[\algorithmicoutput]}

\usepackage{bm}
\usepackage{braket}

\newcommand{\vett}[1]{{\bf{#1}}}
\newcommand{\bts}{\vett{x}}

\newcommand{\nocontentsline}[3]{}
\newcommand{\tocless}[2]{\bgroup\let\addcontentsline=\nocontentsline#1{#2}\egroup}

\title{Protein-Ligand Free Energy Perturbation on Quantum Hardware}
\author{Zhen Li}
\affiliation{Center for Computational Life Sciences, Lerner Research Institute, The Cleveland Clinic, Cleveland, Ohio 44106, United States}
\author{Milana Bazayeva}
\affiliation{Center for Computational Life Sciences, Lerner Research Institute, The Cleveland Clinic, Cleveland, Ohio 44106, United States}
\author{Thaddeus Pellegrini}
\affiliation{IBM Quantum, IBM T.J. Watson Research Center, Yorktown Heights, NY 10598, United States}
\author{Subhamoy Bhowmik}
\affiliation{Center for Computational Life Sciences, Lerner Research Institute, The Cleveland Clinic, Cleveland, Ohio 44106, United States}
\alsoaffiliation{Department of Chemistry, Michigan State University, East Lansing, Michigan 48824, United States}
\author{Susanta Das}
\affiliation{Center for Computational Life Sciences, Lerner Research Institute, The Cleveland Clinic, Cleveland, Ohio 44106, United States}
\author{Danil Kaliakin}
\affiliation{Center for Computational Life Sciences, Lerner Research Institute, The Cleveland Clinic, Cleveland, Ohio 44106, United States}
\author{Fangchun Liang}
\affiliation{Center for Computational Life Sciences, Lerner Research Institute, The Cleveland Clinic, Cleveland, Ohio 44106, United States}
\author{Akhil Shajan}
\affiliation{Center for Computational Life Sciences, Lerner Research Institute, The Cleveland Clinic, Cleveland, Ohio 44106, United States}
\alsoaffiliation{Department of Chemistry, Michigan State University, East Lansing, Michigan 48824, United States}
\author{Kenneth M. Merz Jr.}
\email{kmerz1@gmail.com}
\affiliation{Center for Computational Life Sciences, Lerner Research Institute, The Cleveland Clinic, Cleveland, Ohio 44106, United States}
\alsoaffiliation{Department of Chemistry, Michigan State University, East Lansing, Michigan 48824, United States}

\begin{document}

The use of free energy perturbation (FEP) methods to study protein-ligand complexes is one of the most important tools in structure-based drug design. Because FEP methods typically rely on force fields, they may suffer from force field parameter-related issues. Herein, we present a quantum mechanics/molecular mechanics (QM/MM) hybrid method to overcome deficiencies in force-field models by using QM bookending approaches on both classical and quantum hardware. In the MM part of this QM/MM FEP method, AMBER is used to simulate the protein receptor and the unperturbed moiety of the ligand, with the ff19SB and GAFF2 force fields. In the QM part, QUICK was used to conduct Hartree-Fock (HF) calculations, followed by heat-bath configuration interaction (HCI) as a benchmark on classical devices. To enable the HCI function in QUICK, we developed a Python-based interface to execute HCI from IBM's qiskit-addon-dice-solver. Moreover, the same interface also enabled this work to execute QM/MM FEP calculations on quantum hardware using the Local Unitary Cluster Jastrow (LUCJ) ansatz, followed by sample-based diagonalization (SQD) and extended-SQD (extSQD) post-processing. Using a series of thermolysis inhibitors as an example, we find reasonable agreement with experiment between the classical HCI method and the LUCJ-SQD/extSQD method, with the latter yielding a result closer to the experimental value. The execution time between the HCI-based FEP method and the LUCJ-SQD/extSQD-based FEP method is also comparable, indicating a high potential for utility in the noisy intermediate-scale quantum (NISQ) era.


\section{Introduction}

In drug discovery, a myriad of methods have been applied to accurately predict molecular binding affinities, especially between small-molecule ligands and protein receptors \cite{Schrondinger2015, Schrodinger2023, mmpbsa2012, mmgbsa2013, docking2023}. Among all the methods, the free energy perturbation (FEP) and alchemical free energy (AFE) methods \cite{afe} have been widely accepted as state-of-the-art tools that balance speed and accuracy for calculating relative binding free energies. \cite{devivo2016, jorgensen2000, li2024, saloahen2021}. 
However, while FEP has generally higher accuracy compared to some even faster methods, such as MM/GBSA/PBSA \cite{gbsapbsa2015, gbsapbsa2019, gbsa2022} and docking methods such as Autodock Vina \cite{autodockvina}, such a high accuracy is still based on and limited to the selection and parameterization of force fields (FF). Various recent studies have tackled this challenge using machine-learning force fields \cite{ani1, ani2, ani2024, zeng2023, giese2022} or accounting for polarization effects \cite{clemente2023, Kar2023} in complex protein environments.  

Compared with FEP approaches based on modified force fields, the quantum mechanics/molecular mechanics (QM/MM) hybrid method \cite{warshel1976, duarte2015, Kar2023, manathunga2022, Lu01092016} offers an alternative approach to achieving high accuracy. By carefully constructing the link atoms between the QM and MM regions, the QM/MM method can, in principle, balance simulation speed and accuracy. 

Based on previous work using reference QM/MM potentials \cite{Warshel1992, Gao1992, gao1992priori} and perturbation Hamiltonians \cite{molecules24040681, hudson2018force, hudson2018accelerating, Kearns2017, konig2014, konig2014multiscale, konig2015, konig2018, konig2018comparison, hudson2015}, Giese et al. introduced their book-ending approach for force-field parameterization. \cite{giese2019} In this method, the molecular mechanics (MM) parameters are optimized to reproduce the forces obtained from the QM/MM calculations using a sequence of reference potentials, offering customizable trade-offs between speed and accuracy. However, since the system must be perturbed from the MM force field to the QM Hamiltonian, and then apply corrections to both the MM and QM/MM end states using pure MM free energy methods, the QM/MM FEP method still requires configurational sampling   \cite{giese2024, york2023, Li2003}. One solution is to incorporate a machine-learning (ML)-based reference potential into the book-ending framework to reduce the sampling burden and enhance accuracy. \cite{giese2024} 

Beyond the use of ML tools to construct reference potentials, density functional theory (DFT) is frequently employed as the quantum-mechanical (QM) component in such workflows due to its favorable balance between computational cost and accuracy. Although DFT is fast, it can fail to deliver sufficient accuracy for certain systems.\cite{dftrew} On the other end of the speed-accuracy spectrum, the full configuration interaction (FCI) method provides an exact solution to the CI problem within a chosen basis set by including all possible electron configurations. Consequently, FCI serves as the benchmark for evaluating the performance of less accurate quantum methods.\cite{Gao2024, Vogiatzis2017} Unfortunately, the computational scaling of FCI makes it currently infeasible for large systems, necessitating the use of approximate or heuristic approaches to achieve tractable results.

One such approach is Heat-bath configuration interaction (HCI), which retains only the most energetically significant determinants in order to approximate FCI-level accuracy at a substantially reduced computational cost.\cite{Abraham, Gao2024} However, the classical heuristics used in determinant selection can be inefficient, sometimes compromising accuracy and increasing computational overhead.\cite{sharma2017semistochastic} Because both FCI and HCI are NP-hard problems on classical hardware, there is growing motivation to explore alternative strategies for more efficient traversal of Hilbert space.\cite{holmes2016heat}

Therefore, in a previous work \cite{bazayeva}, we developed a quantum-centric book-ending approach for fast QM/MM alchemical free energy (AFE) calculations using IBM's state-of-the-art quantum hardware. The addition of an accurate HCI-like quantum-centric calculation allows the regular book-ending approach to reduce sampling over fewer $\lambda$ perturbation windows. \cite{giese2024} In short, the HCI-like method is called sample-based quantum diagonalization (SQD) \cite{robledo2024chemistry, shajan2024towards, Liepuoniute2024, yu2025quantum} and its extended combination (SQD-extSQD) \cite{Barison2025}, which leverages statistical sampling from the Hilbert space to overcome the limitations of both FCI and HCI. This enables favorable scaling and achieves good accuracy even for large systems that remain intractable to conventional classical methods.

To ensure efficient communication between the quantum processing unit (QPU) and classical CPUs, we have developed a customized interface. This interface not only facilitates hybrid quantum–classical workflows but also supports conventional HCI simulations, which serve as benchmarks for quantum-centric SQD calculations. 

The interface itself combines the computational efficiency of Fortran 90 and the flexibility of Python 3.11. Fortran underpins high-performance routines in AMBER and the ab initio electronic-structure package QUICK using the library-based application programming interface (API), while Python provides a user-friendly environment for constructing and executing quantum-centric SQD workflows on QPU hardware.

In conclusion, this study establishes a quantum-centric, CI-corrected book-ending framework as a foundational methodology for hybrid quantum–classical simulations. The systems examined do not exhibit strong electron correlation effects~\cite{Raghavachari, Martin}, but they involve a large active space (26 electrons, 19 or 18 orbitals), solvent exchange, and complex ligand–protein interactions, including notable Coulombic and hydrogen-bonding contributions. Furthermore, the thermolysin–inhibitor complexes (PDB IDs: 5TMN and 6TMN)~\cite{matthews1987, merzmkollmanthermolysin} play a critical role in archaeal proteostasis control through their endoproteolytic activity. Collectively, these features highlight the realistic biochemical complexity captured by the present framework and emphasize the promise of quantum hardware in tackling practical challenges in drug discovery.

\section{Methods and Computational Details}

\subsection{Structure preparation and pure MM calculation}

\begin{figure}
     \centering
     \includegraphics[width=385pt]{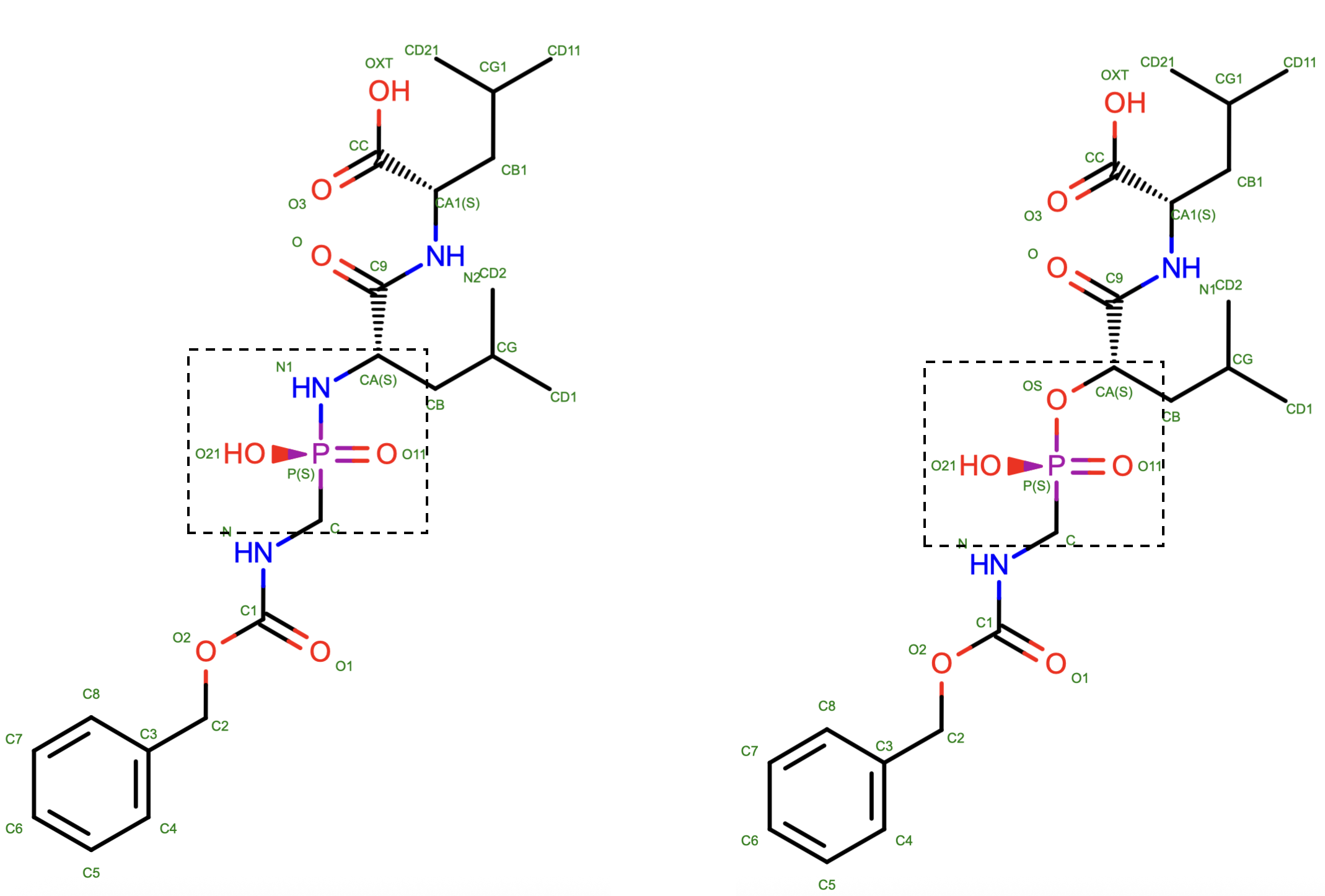}
     \caption{Structures of the two thermolysin inhibitors that differ by one hydrogen bond, with dashed boxes indicating the QM region for the QM/MM simulation. Here, 0PJ (left) forms an extra hydrogen bond with Asn112 of thermolysin using its extra hydrogen on N1, which was not observed in 0PI, since the N1 atom is replaced by an OS atom.}
     \label{fig1}
\end{figure}

The initial coordinates of the thermolysin apo protein were generated using the LEaP module in the AmberTools25 software suite~\cite{amber25}. The thermolysin inhibitors, Cbz-Gly$^P$-(NH)-Leu-Leu (PDB ID: 0PJ) and Cbz-Gly$^P$-(O)-Leu-Leu (PDB ID: 0PI) ~\cite{matthews1987, merzmkollmanthermolysin}, whose chemical structures are shown in (Figure~\ref{fig1}), were parameterized using the General AMBER Force Field (GAFF)~\cite{wang2020gaff2}. Within this procedure, atom types and bonded parameters were automatically assigned by LEaP. Atomic partial charges were obtained using the Restrained Electrostatic Potential (RESP) fitting method~\cite{woods2000resp}, based on molecular electrostatic potentials computed at the B3LYP/6-31+G* level of theory with the Gaussian software package~\cite{g16}.

For each of the two systems (water and protein), the two inhibitors, 0PJ and 0PI, were overlaid, with only the coordination difference at the phosphoramidate (for 0PJ) or phosphate ester (for 0PI) to enable alchemical transformations in subsequent free-energy calculations. The system was then solvated in a cubic water box with at least 24 Å between the solute and the box edges. Periodic boundary conditions (PBC) were applied to eliminate edge effects and emulate a bulk solvent environment. The OPC water model~\cite{onufriev2014opc} was used for all solvated systems, including both inhibitor–water and inhibitor–protein complexes.

Each of the two systems — 0PJ to 0PI mutation in water, and 0PJ to 0PI mutation while bound to the protein — underwent a two-stage energy minimization: first, 10,000 steps of steepest descent, followed by 10,000 steps of conjugate gradient minimization. Electrostatic interactions were treated using the Particle Mesh Ewald (PME) method ~\cite{darden1993pme} with a real-space cutoff of 10 Å and a FFT grid size of 48 × 48 × 48.

After minimization, an NVT equilibration was performed for 360 ps, during which the systems were gradually heated from 0 K to 300 K in 50 K increments. Langevin dynamics with a collision frequency of 2 ps$^{-1}$ was used for temperature control, ensuring stable thermal regulation. Subsequently, an NPT equilibration was conducted at 300 K and 1 atm for 300 ps, with pressure regulation provided by the Berendsen barostat ~\cite{berendsen1984bath} using a reference pressure of 1 bar.

Throughout all stages—from minimization to equilibration and production—the thermodynamic integration (TI) framework was activated via a specific atom mask applied to the phosphoramidate and phosphate ester moieties of each inhibitor, respectively. The fully equilibrated configurations were then used for FEP to obtain the relative binding free energy difference using thermodynamic cycles, as shown in (Figure~\ref{fig2}). A seven-window Thermodynamic Integration (TI) procedure was applied according to the equation given below:

\begin{equation}\label{eq1}
    U(\lambda) = (1-\lambda)U_0 + \lambda U_1
    \;
\end{equation}

In Eq.~\eqref{eq1}, $U(\lambda)$, $U_0$, and $U_1$ denote the overall energy of the perturbed system at a specific $\lambda$ window, the energy of the initial state ($\lambda=0$, which is the molecule dissolved in water), and the energy of the final state ($\lambda=1$, which is the molecule "disappeared" to a dummy molecule in water). The $\Delta A$ between the initial and final system can then be expressed as: 
\begin{equation}\label{eq2}
    \Delta A = \int_{0}^{1} \left\langle \frac{dU}{d\lambda} \right\rangle \,d\lambda
\end{equation}

Because the ensemble energy transition from the initial state to the final state follows Gaussian quadrature rules. We can rewrite Eq.~\eqref{eq2} to
\begin{equation}\label{eq3}
    \Delta A = \sum_{i=1}^{7} \left( c_i \times \left\langle \frac{dU_i}{d\lambda_i} \right\rangle \right)
\end{equation}

In Eq.~\eqref{eq3}, $c_i$ are the Gaussian quadrature coefficients \{0.065, 0.140, 0.191, 0.209, 0.191, 0.140, 0.065\} and $\lambda_i$ are the discrete values \{0.025, 0.130, 0.297, 0.500, 0.703, 0.870, 0.975\} to mix the initial and final values. The entire TI simulation takes 21 ns (3 ns per of the 7 $\lambda$ windows) under NVT to obtain the Helmholtz free energy, which is then converted to the Gibbs free energy in the NPT ensemble. 

\begin{figure}
     \centering
     \includegraphics[width=385pt]{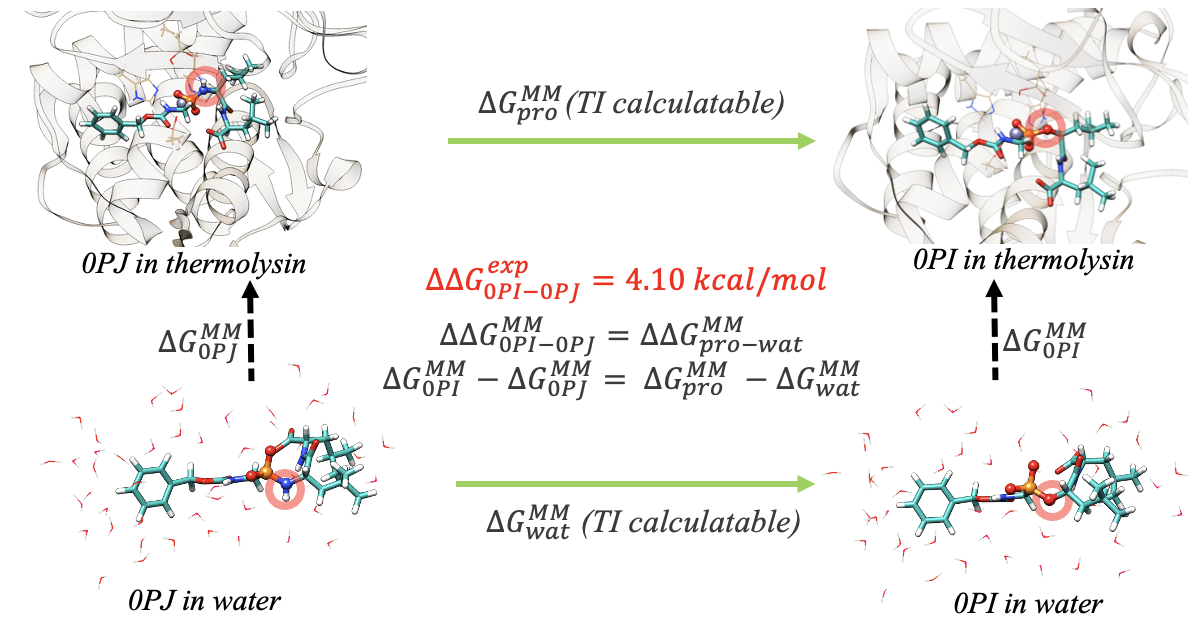}
     \caption{Schematic representation of the pure MM thermodynamic cycle, where green arrows indicate the free energy that FEP can calculate, while black dashed arrows were not computed herein. Using the thermodynamic cycle, these two sets of arrows can be shown to both obtain the relative free energy of binding (see the equations in black).}
     \label{fig2}
\end{figure}

\subsection{QUICK-API-assisted MM-QM/MM perturbation}

\begin{figure}
     \centering
     \includegraphics[width=385pt]{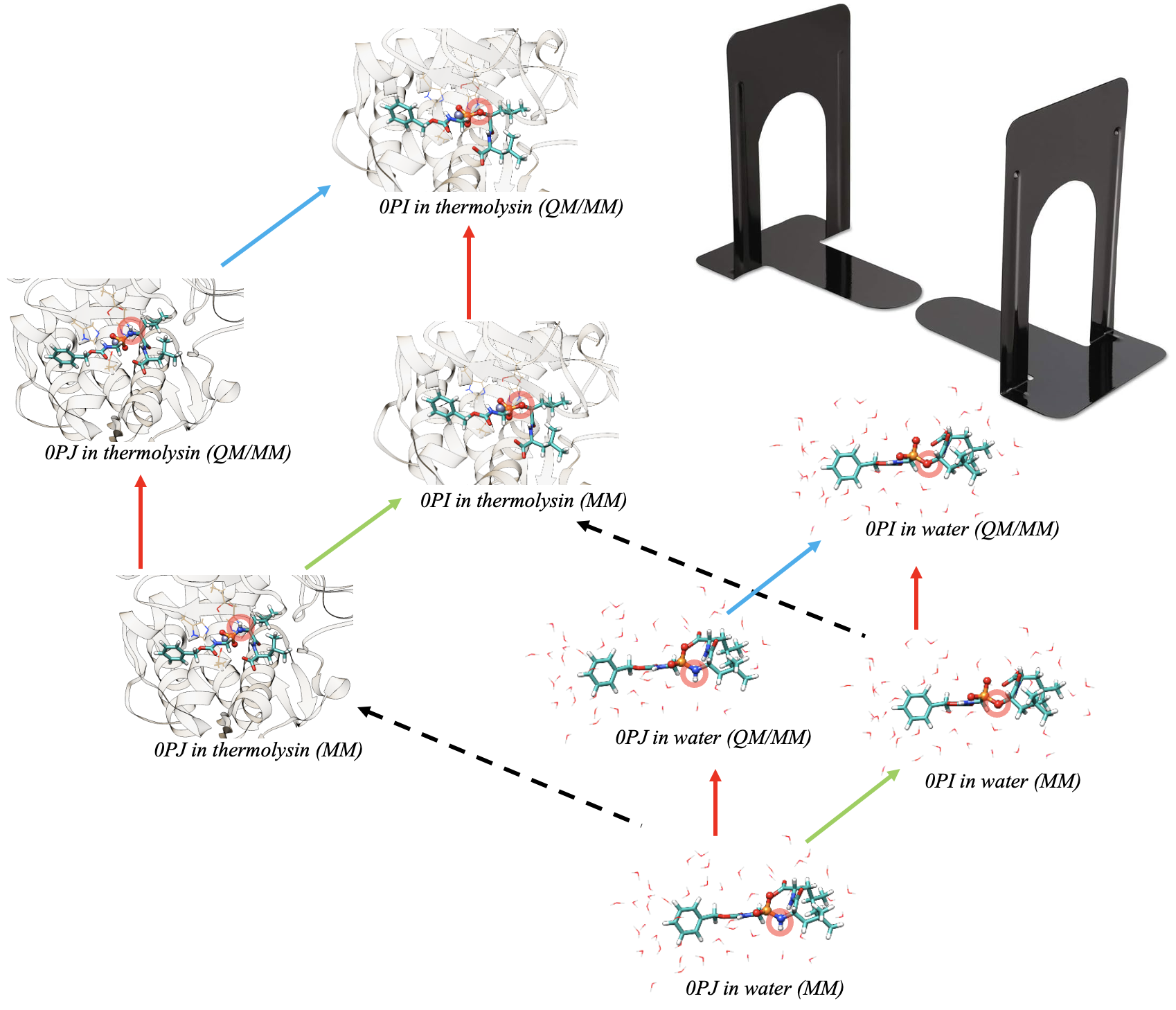}
     \caption{Illustration of how the book-ending correction is applied on the classical MM FEP results. The relative binding free energy is first computed using classical molecular mechanics methods, as shown in (Figure~\ref{fig2}). The book-ending framework then gradually transitions the potential from MM to QM/MM at each end state (vertical red arrows). The quantum mechanical contribution can be evaluated through one of three routes: RHF calculations via the QUICK package, HCI via PySCF and qiskit-addon-dice-solver, or the quantum-centric LUCJ-SQD-extSQD method. The free energy differences are then analyzed using MBAR, producing a quantum correction that is subsequently added to the MM-calculated value (solid blue arrow).}
     \label{fig3}
\end{figure}

After obtaining the pure MM relative binding free energies, the book-ending correction method was then applied to the four end-states of the two systems, as shown in Figure~\ref{fig3}. 

The correction term was obtained by gradually transforming the system from a purely molecular mechanics (MM) potential ($\lambda$ = 0) to a QM/MM representation ($\lambda$ = 1) through a series of intermediate coupling states. The Multistate Bennett Acceptance Ratio (MBAR) method was then applied to evaluate the free energy difference between the two theoretical levels, providing a correction that refines the classically computed relative binding free energy. 

Initial coordinates for 0PJ, 0PI, and the apo thermolysin were prepared following the same protocol as in the classical calculations. For each equilibrated solute, six $\lambda$ windows (0.00, 0.20, 0.40, 0.60, 0.80, 1.00) were simulated under the NVT ensemble with periodic boundary conditions (PBC). Temperature control at 298 K was maintained using Langevin dynamics with a collision frequency of 5 ps$^{-1}$. No pressure coupling was applied, and a 10 Å nonbonded cutoff was used. The SHAKE algorithm constrained all bonds involving hydrogen, except those within the QM-treated solute.

Each book-ending simulation employed two nearly identical input files per $\lambda$ window—one for a classical MM calculation and the other for a QM/MM calculation with electrostatic embedding. The QM region, which consists of the solute molecule, was treated at the Restricted HF (RHF) level using the 6-31+G* base set, since the QM regions (Figure~\ref{fig3}) included the negatively charged phosphate moiety. The entire setup was implemented in the QUICK engine using the API. ~\cite{manathunga2023quantum, cruzeiro2021open}

At predefined intervals (here we used every 10 MD steps), an external CI solver was invoked to compute QM forces, enabling either HCI calculation for the classical benchmark or LUCJ-SQD-extSQD (Figure~\ref{fig3}). Additional details are provided in the Methods section of the Supplementary Information.

Each $\lambda$ window underwent 1 ps equilibration followed by 1 ps of production, where equilibration was performed sequentially across windows, but production runs were initiated independently using the equilibrated structures. Thus, each solute was simulated for a total of 6 ps (1 ps per $\lambda$ window). Simulations were executed under three computational configurations:

(A) AMBER-QUICK RHF only,
(B) AMBER-QUICK, coupled with an HCI solver, and
(C) AMBER-QUICK interfaced with the LUCJ-SQD-extSQD method.

\subsection{Interfacing AMBER with external CI solvers}

\begin{figure}
    \centering
    \includegraphics[width=1\linewidth]{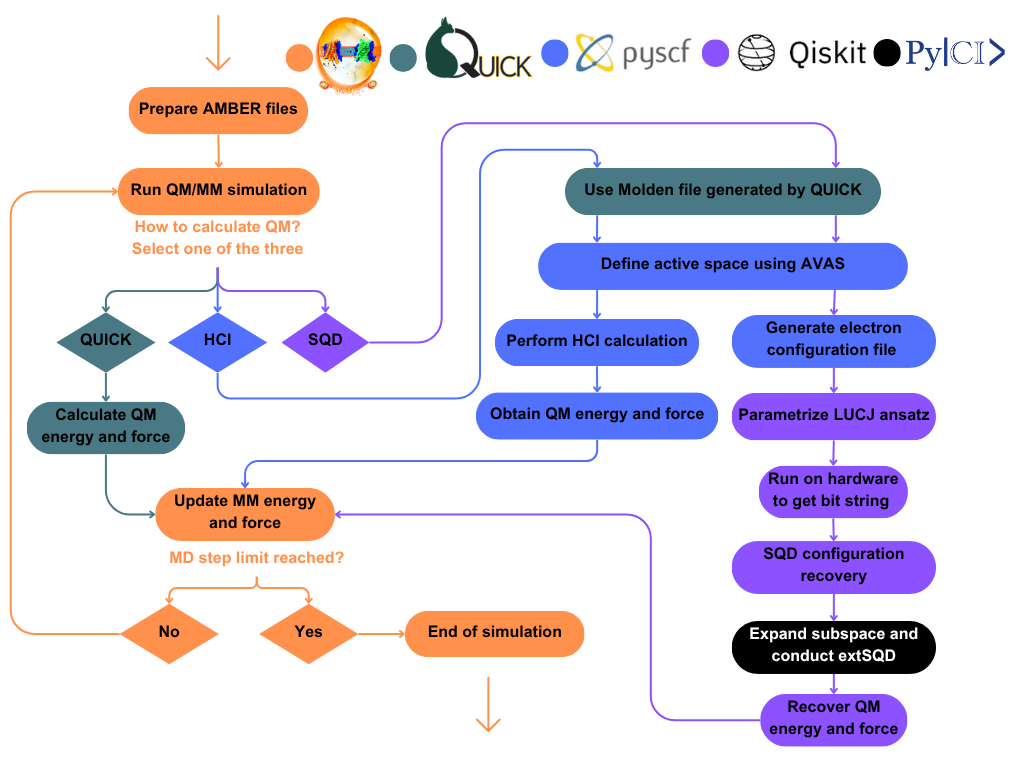}
    \caption{
    Workflow of the standard AMBER/QUICK API functionality within the QM/MM scheme (left), extended with the interface proposed in this work (middle and right). 
    The steps managed by AMBER are shown in orange, those performed by the QUICK engine in green, and HCI simulations are performed using PySCF (blue) and qiskit-addon-dice-solver (purple), while for quantum-centric LUCJ-SQD-extSQD, it starts with the QUICK-generated Molden file, then uses PySCF and a modified, Cartesian-based AVAS \cite{Sayfutyarova2017} to generate the fcidump file for the active space molecular orbitals (MOs), and finally invokes LUCJ and qiskit-addon-sqd (purple) and concludes with PyCI-assisted extSQD (black) configuration recovery.}
    \label{fig4}
\end{figure}

A custom interface was developed to bridge AMBER/QUICK with the Qiskit ecosystem ~\cite{aleksandrowicz2019qiskit, javadi2024quantum}, providing seamless access to quantum hardware and enabling SQD-extSQD post-processing ~\cite{robledo2024chemistry, kaliakin2024accurate, yu2025quantum, Liepuoniute2024, Barison2025, barroca2025surface}. In addition, this same interface was extended to support classical HCI calculations, which were employed to benchmark and validate the accuracy of our quantum-centric results.
(Figure~\ref{fig4}). 

Both CI-based approaches are seamlessly integrated into the existing QUICK API library~\cite{manathunga2023quantum, cruzeiro2021open}, enabling the evaluation of QM energies and gradients using the post-HF method at each simulation step. This design allows CI-level corrections to be periodically applied along the MD trajectory. At user-specified intervals—defined by the CI\_stride keyword in the MD input file—the QM computation is redirected to one of two external solvers:

(A) an HCI solver interfaced through PySCF ~\cite{sun2018pyscf, sun2020recent} and qiskit-addon-dice-solver, or
(B) The SQD-extSQD solver, which utilizes quantum hardware to generate initial electron configurations based on the LUCJ Ansatz ~\cite{motta2023bridging}, followed by post-processing through the SQD-extSQD method ~\cite{robledo2024chemistry, kaliakin2024accurate, yu2025quantum, Liepuoniute2024, Barison2025, barroca2025surface}. 

In the standard workflow (illustrated on the left side of Figure~\ref{fig4}), AMBER/QUICK orchestrates the interaction between the classical and quantum components. At each MD step, AMBER computes all MM contributions, after which the QUICK API library invokes QUICK to evaluate energies and nuclear gradients for the QM region. Once the QUICK calculation completes, the results are then integrated into AMBER through QM link atoms and external point changes. The MD trajectory can then be propagated by computing electrostatic and van der Waals interactions within the MM region and between the QM and MM regions, advancing the simulation in a fully integrated QM/MM loop.

In the extended workflow (the middle and right sides of Figure~\ref{fig4}), the same coupling scheme is retained. QUICK computes RHF energies and gradients at each step, while AMBER integrates QM/MM interactions to propagate dynamics. However, at user-defined CI\_stride intervals, the QM computation is redirected to an external CI solver. This is achieved by leveraging the Molden file generated by QUICK, which contains the geometric and molecular-orbital information required for external CI processing. The energies and gradients returned by the CI solver are then reintegrated into the AMBER framework, allowing MD propagation to resume with updated total energies while maintaining the efficiency and modularity of the original QUICK interface.

\begin{figure}
    \centering
    \includegraphics[width=1\linewidth]{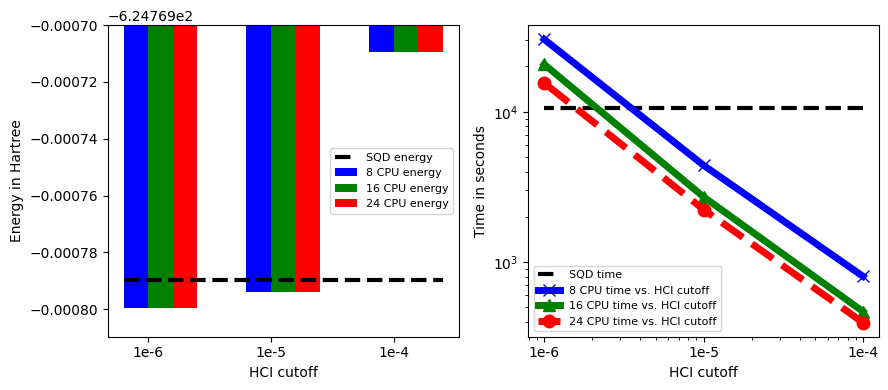}
    \caption{
    Energy and timing benchmarks of HCI with different numbers of CPUs (250GB each) and heat-bath cutoffs, where the horizontal dashed black line indicates the SQD energy and timing, using 10 batches, run with 8 CPUs and 100 GB of memory. Here, the energy and timing contributions from extSQD with a 10$^{-5}$ d-prime single-excitation cutoff (approximately 40 minutes per step) were excluded to match the non-excited HCI setup.}
    \label{fig5}
\end{figure}

Energy and timing benchmarks between the HCI method and LUCJ-SQD are presented in Figure~\ref{fig5}. An ext-HCI \cite{exthci2025} variant would be needed for a better comparison with ext-SQD, and the authors are working to implement ext-HCI for future use. Eventually, an HCI cutoff of 10$^{-5}$ was selected to balance speed and time, since at 8 CPUs, the LUCJ-SQD can reach energy level and timing cost close to the 10$^{-5}$-cutoff HCI. 


\section{Results and Discussion}
\begin{figure}
    \centering
    \includegraphics[width=1\linewidth]{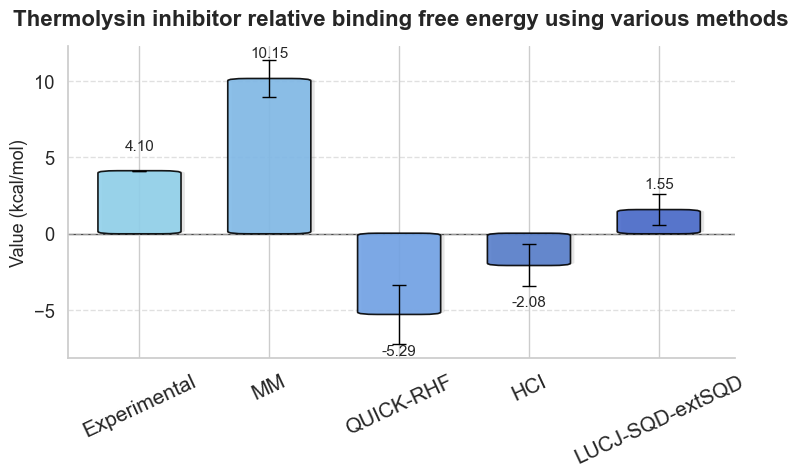}
    \caption{
    Comparison of energy obtained using different methods against the experimental energy, replacing the 0PJ ligand with 0PI for the thermolysin (with a hydrogen bond loss).}
    \label{fig6}
\end{figure}

In this work, we developed an interface to enable the computation of book-ending corrections using HCI, aiming to achieve a higher level of QM accuracy. HCI provides the most accurate QM solution, but its computational cost is high for larger systems, especially when selecting the significant CI vectors and coefficients at a tight heat-bath cutoff. To address this limitation, we also implemented a quantum-centric LUCJ-SQD-extSQD approach. This alternative enables energy evaluation with good accuracy, with potential for further improvements. 
To validate the interface and evaluate its impact on the accuracy of book-ending corrections, we computed the relative binding free energies of 0PJ and 0PI to apo thermolysin, and then applied three quantum-level corrections: QUICK (an RHF calculation), HCI, and LUCJ-SQD-extSQD. All QM calculations were performed with the 6-31+G* basis set, as stated in the method section. Figure~\ref{fig6} presents a comprehensive overview, including the experimental relative binding free energy, the MM-calculated value, and the book-ending corrected value. \cite{matthews1987, merzmkollmanthermolysin} As is evident from Figure~\ref{fig6}, classical MM calculations exhibit significant deviations from the experimental value by 6.05 kcal/mol, yet the QUICK RHF calculation has the largest deviation of -9.39 kcal/mol, showing that without the addition of post-HF approaches, RHF calculations appear to over-correct the relative binding free energy. Interestingly, the introduction of HCI and LUCJ-SQD-extSQD improves the outcome in this case, with HCI showing a deviation of 6.18 kcal/mol and LUCJ-SQD-extSQD showing a deviation of 2.54 kcal/mol. Ideally, the LUCJ-SQD-extSQD result can be further improved by increasing the number of samples in the SQD-extSQD method, or by running LUCJ iteratively \cite{shirakawalucj2025}. 

For the first time in protein systems, we demonstrate the integration of quantum hardware into classical alchemical workflows used in drug discovery, enabling enhanced accuracy in free energy predictions through the inclusion of both the HCI solver and the LUCJ-SQD-extSQD workflow, where the LUCJ-SQD-extSQD method yields slightly lower energy yet takes slightly longer time to execute than the HCI method. This achievement represents a significant milestone in advancing hybrid quantum–classical simulations that can be used in drug discovery campaigns.

Figure~\ref{fig6} reveals consistent and reproducible corrections across the three quantum mechanical approaches examined. The observed numerical variations among the RHF, HCI, and LUCJ–SQD–extSQD methods reflect their fundamental methodological differences rather than differences in their implementations.

Collectively, these findings validate the hybrid QM/MM interface developed in this study, which improves relative binding free energy predictions by incorporating multiple quantum backends within a unified computational framework. Furthermore, the interface is designed to be modular and extensible, allowing dynamic switching between classical and advanced quantum solvers. Importantly, this work extends the previous successful embedding of book-ending calculation using quantum hardware, establishing a foundation for future quantum-enhanced molecular simulations for drug discovery.

\section{Conclusions and Outlook}

FEP is widely used to estimate free energy differences associated with molecular transformations, providing crucial insights into properties such as binding affinities and reaction energetics. Although quantum algorithms have been applied to general electronic structure problems, their integration into FEP workflows remains largely unexplored. In this study, we introduce quantum computing into protein-ligand FEP calculations, addressing a critical yet underdeveloped area with significant potential applications in drug discovery and computational chemistry.

Quantum algorithms tailored for free energy estimation represent a promising frontier for applying quantum hardware to real-world chemical challenges. A major contribution of this work is the integration of CI-like methods, specifically HCI and SQD-extSQD, into established classical frameworks. Historically, book-ending corrections have relied solely on RHF and DFT methods; the present work extends this paradigm by incorporating CI-level quantum corrections on both energy and force. 

Beyond improved accuracy, FEP frameworks are naturally suited for parallelization across multiple quantum devices, since the $\lambda$-windows of the production step are independent by design. Leveraging this intrinsic parallelism can significantly enhance the throughput and scalability of quantum simulations, laying the groundwork for next-generation hybrid pipelines optimized for the post-NISQ era. This advancement represents an important step toward scalable, quantum-assisted free-energy predictions for drug discovery and lead-optimization applications.

\begin{acknowledgement}

The authors gratefully acknowledge financial support from the National Science Foundation (NSF) through CSSI Frameworks Grant OAC-2435622 and from the National Institutes of Health (Grant Numbers GM130641). The authors also thank Javier Robledo Moreno, Caleb Johnson, Abdullah Ash Saki, Iskandar Sitdikov, Kevin Sung, and Antonio Mezzacapo for their guidance on the LUCJ ansatz and the SQD method. 

Dr. Mario Motta (IBM T.J. Watson Research Center) kindly provided great help with manuscript review and the AVAS orbital-selection code to support Cartesian coordinates. Dr. Timothy Giese and Dr. Darrin M. York (Rutgers University) provided significant help on the book-ending hands-on tutorial in the early phase of this project. Vikrant Tripathy and Dr. Andreas Goetz (UCSD) provided helpful suggestions on modifying the QUICK API to communicate with CI solvers. The authors hereby deliver special thanks to these colleagues. 

\end{acknowledgement}

\section*{Code Availability}

Qiskit, ffsim, and Qiskit IBM Runtime utilized for LUCJ simulations can be obtained from the corresponding GitHub repositories:

\url{https://github.com/qiskit-community/ffsim}

\url{https://github.com/Qiskit/qiskit}

\url{https://github.com/Qiskit/qiskit-ibm-runtime}

Configuration recovery code is distributed as the Qiskit SQD addon:

\url{https://github.com/Qiskit/qiskit-addon-sqd}

PySCF through the corresponding GitHub repository:

\url{https://github.com/pyscf/pyscf}

The tutorial demonstrating the full SQD workflow is available below:

\url{qiskit.github.io/qiskit-addon-sqd/tutorials/01_chemistry_hamiltonian.html}

AMBER/QUICK API interface for free energy calculation can be obtained upon request.

\section*{Competing Interest}
The authors declare no competing interests.


\bibliography{AFE_QPU}

@article{docking2023,
   author = {Agu, P. C. and Afiukwa, C. A. and Orji, O. U. and Ezeh, E. M. and Ofoke, I. H. and Ogbu, C. O. and Ugwuja, E. I. and Aja, P. M.},
   title = {Molecular docking as a tool for the discovery of molecular targets of nutraceuticals in diseases management},
   journal = {Scientific Reports},
   volume = {13},
   number = {1},
   pages = {13398},
   ISSN = {2045-2322},
   DOI = {10.1038/s41598-023-40160-2},
   url = {https://doi.org/10.1038/s41598-023-40160-2},
   year = {2023},
   type = {Journal Article}
}

@article{autodockvina,
author = {Eberhardt, Jerome and Santos-Martins, Diogo and Tillack, Andreas F. and Forli, Stefano},
title = {AutoDock Vina 1.2.0: New Docking Methods, Expanded Force Field, and Python Bindings},
journal = {Journal of Chemical Information and Modeling},
volume = {61},
number = {8},
pages = {3891-3898},
year = {2021},
doi = {10.1021/acs.jcim.1c00203},
    note ={PMID: 34278794},
URL = {https://doi.org/10.1021/acs.jcim.1c00203},
eprint = {https://doi.org/10.1021/acs.jcim.1c00203}
}

@article{mmpbsa2012,
author = {Miller, Bill R. III and McGee, T. Dwight Jr. and Swails, Jason M. and Homeyer, Nadine and Gohlke, Holger and Roitberg, Adrian E.},
title = {MMPBSA.py: An Efficient Program for End-State Free Energy Calculations},
journal = {Journal of Chemical Theory and Computation},
volume = {8},
number = {9},
pages = {3314-3321},
year = {2012},
doi = {10.1021/ct300418h},
    note ={PMID: 26605738},
URL = {https://doi.org/10.1021/ct300418h},
eprint = {https://doi.org/10.1021/ct300418h}
}

@article{mmgbsa2013,
author = {Greenidge, Paulette A. and Kramer, Christian and Mozziconacci, Jean-Christophe and Wolf, Romain M.},
title = {MM/GBSA Binding Energy Prediction on the PDBbind Data Set: Successes, Failures, and Directions for Further Improvement},
journal = {Journal of Chemical Information and Modeling},
volume = {53},
number = {1},
pages = {201-209},
year = {2013},
doi = {10.1021/ci300425v},
    note ={PMID: 23268595},
URL = {https://doi.org/10.1021/ci300425v},
eprint = {https://doi.org/10.1021/ci300425v}
}

@article{gbsapbsa2015,
   author = {Genheden, S. and Ryde, U.},
   title = {The MM/PBSA and MM/GBSA methods to estimate ligand-binding affinities},
   journal = {Expert Opin Drug Discov},
   volume = {10},
   number = {5},
   pages = {449-61},
   ISSN = {1746-0441 (Print) 1746-0441},
   DOI = {10.1517/17460441.2015.1032936},
   year = {2015},
   type = {Journal Article}
}

@article{gbsapbsa2019,
author = {Wang, Ercheng and Sun, Huiyong and Wang, Junmei and Wang, Zhe and Liu, Hui and Zhang, John Z. H. and Hou, Tingjun},
title = {End-Point Binding Free Energy Calculation with MM/PBSA and MM/GBSA: Strategies and Applications in Drug Design},
journal = {Chemical Reviews},
volume = {119},
number = {16},
pages = {9478-9508},
year = {2019},
doi = {10.1021/acs.chemrev.9b00055},
    note ={PMID: 31244000},
URL = {https://doi.org/10.1021/acs.chemrev.9b00055},
eprint = {https://doi.org/10.1021/acs.chemrev.9b00055}
}

@article{gbsa2022,
   author = {Dasmahapatra, U. and Kumar, C. K. and Das, S. and Subramanian, P. T. and Murali, P. and Isaac, A. E. and Ramanathan, K. and Mm, B. and Chanda, K.},
   title = {In-silico molecular modelling, MM/GBSA binding free energy and molecular dynamics simulation study of novel pyrido fused imidazo[4,5-c]quinolines as potential anti-tumor agents},
   journal = {Front Chem},
   volume = {10},
   pages = {991369},
   ISSN = {2296-2646 (Print)
2296-2646},
   DOI = {10.3389/fchem.2022.991369},
   year = {2022},
   type = {Journal Article}
}

@article{Schrondinger2015,
author = {Wang, Lingle and Wu, Yujie and Deng, Yuqing and Kim, Byungchan and Pierce, Levi and Krilov, Goran and Lupyan, Dmitry and Robinson, Shaughnessy and Dahlgren, Markus K. and Greenwood, Jeremy and Romero, Donna L. and Masse, Craig and Knight, Jennifer L. and Steinbrecher, Thomas and Beuming, Thijs and Damm, Wolfgang and Harder, Ed and Sherman, Woody and Brewer, Mark and Wester, Ron and Murcko, Mark and Frye, Leah and Farid, Ramy and Lin, Teng and Mobley, David L. and Jorgensen, William L. and Berne, Bruce J. and Friesner, Richard A. and Abel, Robert},
title = {Accurate and Reliable Prediction of Relative Ligand Binding Potency in Prospective Drug Discovery by Way of a Modern Free-Energy Calculation Protocol and Force Field},
journal = {Journal of the American Chemical Society},
volume = {137},
number = {7},
pages = {2695-2703},
year = {2015},
doi = {10.1021/ja512751q},
    note ={PMID: 25625324},
URL = {https://doi.org/10.1021/ja512751q},
eprint = {https://doi.org/10.1021/ja512751q}
}

@article{Schrodinger2023,
   author = {Ross, Gregory A. and Lu, Chao and Scarabelli, Guido and Albanese, Steven K. and Houang, Evelyne and Abel, Robert and Harder, Edward D. and Wang, Lingle},
   title = {The maximal and current accuracy of rigorous protein-ligand binding free energy calculations},
   journal = {Communications Chemistry},
   volume = {6},
   number = {1},
   pages = {222},
   ISSN = {2399-3669},
   DOI = {10.1038/s42004-023-01019-9},
   url = {https://doi.org/10.1038/s42004-023-01019-9},
   year = {2023},
   type = {Journal Article}
}

@misc{bazayeva,
   author = {Bazayeva, Milana and Li, Zhen and Kaliakin, Danil and Liang, Fangchun and Shajan, Akhil and Das, Susanta and Merz, Kenneth M., Jr},
   title = {Quantum-Centric Alchemical Free Energy Calculations},
   pages = {arXiv:2506.20825},
   month = {June 01, 2025},
   DOI = {10.48550/arXiv.2506.20825},
   url = {https://ui.adsabs.harvard.edu/abs/2025arXiv250620825B},
   year = {2025},
   type = {Electronic Article}
}

@misc{Liepuoniute2024,
  author       = {Liepuoniute, I. and Doney, K. D. and Robledo-Moreno, J. and Job, J. A. and Friend, W. S. and Jones, G. O.},
  title        = {Quantum-Centric Study of Methylene Singlet and Triplet States},
  year         = {2024},
  eprint       = {2411.04827},
  archivePrefix = {arXiv},
  primaryClass = {physics.chem-ph}
}

@article{Libcint15,
author = {Sun, Qiming},
title = {Libcint: An efficient general integral library for Gaussian basis functions},
journal = {Journal of Computational Chemistry},
volume = {36},
number = {22},
pages = {1664-1671},
doi = {https://doi.org/10.1002/jcc.23981},
url = {https://onlinelibrary.wiley.com/doi/abs/10.1002/jcc.23981},
eprint = {https://onlinelibrary.wiley.com/doi/pdf/10.1002/jcc.23981},
year = {2015}}

@article{barroca2025surface,
  title = {Surface Reaction Simulations for Battery Materials through Sample-Based Quantum Diagonalization and Local Embedding},
  author = {Marco Antonio Barroca and Tanvi Gujarati and Vidushi Sharma and Rodrigo Neumann Barros Ferreira and Young-Hye Na and Maxwell Giammona and Antonio Mezzacapo and Benjamin Wunsch and Mathias Steiner},
  journal = {arXiv preprint arXiv:2503.10923},
  year = {2025},
  note = {Submitted on 13 Mar 2025},
  url = {https://doi.org/10.48550/arXiv.2503.10923},
  eprint = {2503.10923},
  archivePrefix = {arXiv}
}

@article{yu2025quantum,
  title = {Quantum-Centric Algorithm for Sample-Based Krylov Diagonalization},
  author = {Jeffery Yu and Javier Robledo Moreno and Joseph T. Iosue and Luke Bertels and Daniel Claudino and Bryce Fuller and Peter Groszkowski and Travis S. Humble and Petar Jurcevic and William Kirby and Thomas A. Maier and Mario Motta and Bibek Pokharel and Alireza Seif and Amir Shehata and Kevin J. Sung and Minh C. Tran and Vinay Tripathi and Antonio Mezzacapo and Kunal Sharma},
  journal = {arXiv preprint arXiv:2501.09702},
  year = {2025},
  note = {Submitted on 16 Jan 2025 (v1), last revised 24 Jan 2025 (this version, v2)},
  url = {https://doi.org/10.48550/arXiv.2501.09702},
  eprint = {2501.09702},
  archivePrefix = {arXiv},
  primaryClass = {quant-ph}
}

@article{Coe2023,
doi = {10.1021/acs.jctc.2c01062},
author = {Coe, Jeremy P.},
title = {Analytic Gradients for Selected Configuration Interaction},
journal = {Journal of Chemical Theory and Computation},
volume = {19},
number = {3},
pages = {874-886},
year = {2023},
doi = {10.1021/acs.jctc.2c01062},
note ={PMID: 36656261},
URL = {https://doi.org/10.1021/acs.jctc.2c01062},
eprint = {https://doi.org/10.1021/acs.jctc.2c01062}}

@article{Li2003,
  author  = {Li, G. and Zhang, X. and Cui, Q.},
  title   = {Free energy perturbation calculations with combined QM/MM potentials: complications, simplifications, and applications to redox potential calculations},
  journal = {J. Phys. Chem. B},
  year    = {2003},
  volume  = {107},
  pages   = {8643--8653}}

@Inbook{Hunt2023,
author="Hunt, John",
title="Monkey Patching",
bookTitle="A Beginners Guide to Python 3 Programming",
year="2023",
publisher="Springer International Publishing",
address="Cham",
pages="487--490",
isbn="978-3-031-35122-8",
doi="10.1007/978-3-031-35122-8_43",
url="https://doi.org/10.1007/978-3-031-35122-8_43"
}

@article{manathunga2023quantum,
  author    = {Manathunga, M. and Aktulga, H. M. and Götz, A. W. and Merz, K. M.},
  title     = {Quantum Mechanics/Molecular Mechanics Simulations on NVIDIA and AMD Graphics Processing Units},
  journal   = {J. Chem. Inf. Model.},
  year      = {2023},
  volume    = {63},
  pages     = {711--717}}

@article{cruzeiro2021open,
  author    = {Cruzeiro, V. W. D. and Manathunga, M. and Merz, K. M. and Götz, A. W.},
  title     = {Open-Source Multi-GPU-Accelerated QM/MM Simulations with AMBER and QUICK},
  journal   = {J. Chem. Inf. Model.},
  year      = {2021},
  volume    = {61},
  pages     = {2109--2115}}

@article{Barison2025,
  author  = {Barison, S. and Robledo Moreno, J. and Motta, M.},
  title   = {Quantum-centric computation of molecular excited states with extended sample-based quantum diagonalization},
  journal = {Quantum Science and Technology},
  year    = {2025},
  volume  = {10},
  pages   = {025034},
  doi     = {10.1088/2058-9565/ad38c4}}

@article{Raghavachari,
doi = {10.1021/jp953749i},
author = {Raghavachari, Krishnan and Anderson, James B.},
title = {Electron Correlation Effects in Molecules},
journal = {The Journal of Physical Chemistry},
volume = {100},
number = {31},
pages = {12960-12973},
year = {1996},
doi = {10.1021/jp953749i},
URL = {https://doi.org/10.1021/jp953749i},
eprint = {https://doi.org/10.1021/jp953749i}}

@article{Martin,
author = {Martin, Jan M. L.},
title = {Electron Correlation: Nature's Weird and Wonderful Chemical Glue},
journal = {Israel Journal of Chemistry},
volume = {62},
number = {1-2},
pages = {e202100111},
doi = {https://doi.org/10.1002/ijch.202100111},
url = {https://onlinelibrary.wiley.com/doi/abs/10.1002/ijch.202100111},
eprint = {https://onlinelibrary.wiley.com/doi/pdf/10.1002/ijch.202100111},
year = {2022}}

@article{Lu01092016,
author = {Xiya Lu and Dong Fang and Shingo Ito and Yuko Okamoto and Victor Ovchinnikov and Qiang Cui and},
title = {QM/MM free energy simulations: recent progress and challenges},
journal = {Molecular Simulation},
volume = {42},
number = {13},
pages = {1056--1078},
year = {2016},
publisher = {Taylor \& Francis},
doi = {10.1080/08927022.2015.1132317},
note ={PMID: 27563170},
URL = { https://doi.org/10.1080/08927022.2015.1132317},
eprint = {  https://doi.org/10.1080/08927022.2015.1132317}}

@article{clemente2023,
doi = {10.1021/acs.jcim.2c01522},
author = {Clemente, Camila M. and Capece, Luciana and Martí, Marcelo A.},
title = {Best Practices on QM/MM Simulations of Biological Systems},
journal = {Journal of Chemical Information and Modeling},
volume = {63},
number = {9},
pages = {2609-2627},
year = {2023},
doi = {10.1021/acs.jcim.2c01522},
    note ={PMID: 37100031},
URL = {https://doi.org/10.1021/acs.jcim.2c01522},
eprint = {https://doi.org/10.1021/acs.jcim.2c01522}}

@article{dftrew,
doi = {10.1021/cr200107z},
author = {Cohen, Aron J. and Mori-Sánchez, Paula and Yang, Weitao},
title = {Challenges for Density Functional Theory},
journal = {Chemical Reviews},
volume = {112},
number = {1},
pages = {289-320},
year = {2012},
doi = {10.1021/cr200107z},
note ={PMID: 22191548},
URL = { https://doi.org/10.1021/cr200107z},
eprint = { https://doi.org/10.1021/cr200107z}}

@article{Abraham, 
doi = {10.1021/acs.jctc.0c00141},
author = {Abraham, Vibin and Mayhall, Nicholas J.},
title = {Selected Configuration Interaction in a Basis of Cluster State Tensor Products},
journal = {Journal of Chemical Theory and Computation},
volume = {16},
number = {10},
pages = {6098-6113},
year = {2020},
doi = {10.1021/acs.jctc.0c00141},
    note ={PMID: 32846094},
URL = {https://doi.org/10.1021/acs.jctc.0c00141}}

@article{york2023,
doi = {10.1021/acsphyschemau.3c00033},
author = {York, Darrin M.},
title = {Modern Alchemical Free Energy Methods for Drug Discovery Explained},
journal = {ACS Physical Chemistry Au},
volume = {3},
number = {6},
pages = {478-491},
year = {2023},
doi = {10.1021/acsphyschemau.3c00033},
URL = {https://doi.org/10.1021/acsphyschemau.3c00033}}

@article{duarte2015,
title = {Recent advances in QM/MM free energy calculations using reference potentials},
journal = {Biochimica et Biophysica Acta (BBA) - General Subjects},
volume = {1850},
number = {5},
pages = {954-965},
year = {2015},
note = {Recent developments of molecular dynamics},
issn = {0304-4165},
doi = {https://doi.org/10.1016/j.bbagen.2014.07.008},
url = {https://www.sciencedirect.com/science/article/pii/S0304416514002463},
author = {Fernanda Duarte and Beat A. Amrein and David Blaha-Nelson and Shina C.L. Kamerlin}
}

@article{Kar2023,
title = {Benefits of hybrid QM/MM over traditional classical mechanics in pharmaceutical systems},
journal = {Drug Discovery Today},
volume = {28},
number = {1},
pages = {103374},
year = {2023},
issn = {1359-6446},
doi = {https://doi.org/10.1016/j.drudis.2022.103374},
url = {https://www.sciencedirect.com/science/article/pii/S1359644622003671},
author = {Rajiv K. Kar}}

@article{molecules24040681,
author = {Kearns, Fiona L. and Warrensford, Luke and Boresch, Stefan and Woodcock, H. Lee},
title = {The Good, the Bad, and the Ugly: “HiPen”, a New Dataset for Validating (S)QM/MM Free Energy Simulations},
journal = {Molecules},
volume = {24},
year = {2019},
number = {4},
URL = {https://www.mdpi.com/1420-3049/24/4/681},
PubMedID = {30769826},
doi = {10.3390/molecules24040681}
}

@article{manathunga2022,
title = {Computer-aided drug design, quantum-mechanical methods for biological problems},
journal = {Current Opinion in Structural Biology},
volume = {75},
pages = {102417},
year = {2022},
issn = {0959-440X},
doi = {https://doi.org/10.1016/j.sbi.2022.102417},
url = {https://www.sciencedirect.com/science/article/pii/S0959440X22000963},
author = {Madushanka Manathunga and Andreas W. Götz and Kenneth M. Merz}}

@article{devivo2016,
author = {De Vivo, Marco and Masetti, Matteo and Bottegoni, Giovanni and Cavalli, Andrea},
title = {Role of Molecular Dynamics and Related Methods in Drug Discovery},
journal = {Journal of Medicinal Chemistry},
volume = {59},
number = {9},
pages = {4035-4061},
year = {2016},
doi = {10.1021/acs.jmedchem.5b01684},
    note ={PMID: 26807648},
URL = { 
        https://doi.org/10.1021/acs.jmedchem.5b01684
},
eprint = {    
        https://doi.org/10.1021/acs.jmedchem.5b01684
}
}

@article{mbar,
    author = {Shirts, Michael R. and Chodera, John D.},
    title = {Statistically optimal analysis of samples from multiple equilibrium states},
    journal = {The Journal of Chemical Physics},
    volume = {129},
    number = {12},
    pages = {124105},
    year = {2008},
    issn = {0021-9606},
    doi = {10.1063/1.2978177},
    url = {https://doi.org/10.1063/1.2978177},
    eprint = {https://pubs.aip.org/aip/jcp/article-pdf/doi/10.1063/1.2978177/15418484/124105\_1\_online.pdf},
}

@article{jorgensen2000,
title = {Prediction of drug solubility from Monte Carlo simulations},
journal = {Bioorganic $\&$ Medicinal Chemistry Letters},
volume = {10},
number = {11},
pages = {1155-1158},
year = {2000},
issn = {0960-894X},
doi = {https://doi.org/10.1016/S0960-894X(00)00172-4},
url = {https://www.sciencedirect.com/science/article/pii/S0960894X00001724},
author = {William L. Jorgensen and Erin M. Duffy}
}

@article{li2024,
   title={A hybrid quantum computing pipeline for real world drug discovery},
   volume={14},
   ISSN={2045-2322},
   url={http://dx.doi.org/10.1038/s41598-024-67897-8},
   DOI={10.1038/s41598-024-67897-8},
   number={1},
   journal={Scientific Reports},
   publisher={Springer Science and Business Media LLC},
   author={Li, Weitang and Yin, Zhi and Li, Xiaoran and Ma, Dongqiang and Yi, Shuang and Zhang, Zhenxing and Zou, Chenji and Bu, Kunliang and Dai, Maochun and Yue, Jie and Chen, Yuzong and Zhang, Xiaojin and Zhang, Shengyu},
   year={2024},
   month=jul }

@article{saloahen2021,
author={Salo-Ahen, Outi M. H. and Alanko, Ida and Bhadane, Rajendra and Bonvin, Alexandre M. J. J. and Honorato, Rodrigo Vargas and Hossain, Shakhawath and Juffer, André H. and Kabedev, Aleksei and Lahtela-Kakkonen, Maija and Larsen, Anders Støttrup and Lescrinier, Eveline and Marimuthu, Parthiban and Mirza, Muhammad Usman and Mustafa, Ghulam and Nunes-Alves, Ariane and Pantsar, Tatu and Saadabadi, Atefeh and Singaravelu, Kalaimathy and Vanmeert, Michiel},
title={Molecular Dynamics Simulations in Drug Discovery and Pharmaceutical Development},
journal={Processes},
volume={9},
year={2021},
number={1},
ARTICLE-NUMBER={71},
url = {https://www.mdpi.com/2227-9717/9/1/71},
issn = {2227-9717},
doi = {10.3390/pr9010071}
}

@article{warshel1976,
title = {Theoretical studies of enzymic reactions: Dielectric, electrostatic and steric stabilization of the carbonium ion in the reaction of lysozyme},
journal = {Journal of Molecular Biology},
volume = {103},
number = {2},
pages = {227-249},
year = {1976},
issn = {0022-2836},
doi = {https://doi.org/10.1016/0022-2836(76)90311-9},
url = {https://www.sciencedirect.com/science/article/pii/0022283676903119},
author = {A. Warshel and M. Levitt}
}

@article{giese2024,
author = {Giese, Timothy J. and Zeng, Jinzhe and Lerew, Lauren and McCarthy, Erika and Tao, Yujun and Ekesan, {\c{S}\"olen} and York, Darrin M.},
title = {Software Infrastructure for Next-Generation QM/MM-$\triangle$MLP Force Fields},
journal = {The Journal of Physical Chemistry B},
volume = {128},
number = {26},
pages = {6257-6271},
year = {2024},
doi = {10.1021/acs.jpcb.4c01466},
note ={PMID: 38905451},
URL = {https://doi.org/10.1021/acs.jpcb.4c01466},
eprint = {https://doi.org/10.1021/acs.jpcb.4c01466}
}

@article{zeng2023,
author = {Zeng, Jinzhe and Tao, Yujun and Giese, Timothy J. and York, Darrin M.},
title = {QD$\pi$: A Quantum Deep Potential Interaction Model for Drug Discovery},
journal = {Journal of Chemical Theory and Computation},
volume = {19},
number = {4},
pages = {1261-1275},
year = {2023},
doi = {10.1021/acs.jctc.2c01172},
note ={PMID: 36696673},
URL = {https://doi.org/10.1021/acs.jctc.2c01172},
eprint = {https://doi.org/10.1021/acs.jctc.2c01172}
}

@article{giese2022,
author = {Giese, Timothy J. and Zeng, Jinzhe and Ekesan, {\c{S}\"olen} and York, Darrin M.},
title = {Combined QM/MM, Machine Learning Path Integral Approach to Compute Free Energy Profiles and Kinetic Isotope Effects in RNA Cleavage Reactions},
journal = {Journal of Chemical Theory and Computation},
volume = {18},
number = {7},
pages = {4304-4317},
year = {2022},
doi = {10.1021/acs.jctc.2c00151},
note ={PMID: 35709391},
URL = {https://doi.org/10.1021/acs.jctc.2c00151},
eprint = {https://doi.org/10.1021/acs.jctc.2c00151}
}

@article{giese2019,
author = {Giese, Timothy J. and York, Darrin M.},
title = {Development of a Robust Indirect Approach for MM → QM Free Energy Calculations That Combines Force-Matched Reference Potential and Bennett’s Acceptance Ratio Methods},
journal = {Journal of Chemical Theory and Computation},
volume = {15},
number = {10},
pages = {5543-5562},
year = {2019},
doi = {10.1021/acs.jctc.9b00401},
note ={PMID: 31507179},
URL = {https://doi.org/10.1021/acs.jctc.9b00401},
eprint =  https://doi.org/10.1021/acs.jctc.9b00401}

@article{amber25,
author = {Case, David A. and Cerutti, David S. and Cruzeiro, Vinícius Wilian D. and Darden, Thomas A. and Duke, Robert E. and Ghazimirsaeed, Mahdieh and Giambaşu, George M. and Giese, Timothy J. and G{\"o}tz, Andreas W. and Harris, Julie A. and Kasavajhala, Koushik and Lee, Tai-Sung and Li, Zhen and Lin, Charles and Liu, Jian and Miao, Yinglong and Salomon-Ferrrer, Romelia and Shen, Jana and Snyder, Ryan and Swails, Jason and Walker, Ross C. and Wang, Jinan and Wu, Xiongwu and Zeng, Jinzhe and Cheatham III, Thomas E. and Roe, Daniel R. and Roitberg, Adrian and Simmerling, Carlos and York, Darrin M. and Nagan, Maria C. and Merz, Kenneth M. Jr.},
title = {Recent Developments in Amber Biomolecular Simulations},
journal = {Journal of Chemical Information and Modeling},
volume = {65},
number = {15},
pages = {7835-7843},
year = {2025},
doi = {10.1021/acs.jcim.5c01063},
    note ={PMID: 40728386},
URL = {https://doi.org/10.1021/acs.jcim.5c01063},
eprint = {https://doi.org/10.1021/acs.jcim.5c01063}
}

@article{afe,
  author    = {Straatsma, T. P. and McCammon, J. A.},
  title     = {Computational Alchemy},
  journal   = {Annual Review of Physical Chemistry},
  volume    = {43},
  pages     = {407-435},
  year      = {1992},
  doi       = {10.1146/annurev.pc.43.100192.002203}
}

@article{matthews1987,
author = {Dale E. Tronrud  and Hazel M. Holden  and Brian W. Matthews },
title = {Structures of Two Thermolysin-Inhibitor Complexes That Differ by a Single Hydrogen Bond},
journal = {Science},
volume = {235},
number = {4788},
pages = {571-574},
year = {1987},
doi = {10.1126/science.3810156},
URL = {https://www.science.org/doi/abs/10.1126/science.3810156},
eprint = {https://www.science.org/doi/pdf/10.1126/science.3810156},
}

@article{merzmkollmanthermolysin,
   author = {Merz, Kenneth M., Jr. and Kollman, Peter A.},
   title = {Free energy perturbation simulations of the inhibition of thermolysin: prediction of the free energy of binding of a new inhibitor},
   journal = {Journal of the American Chemical Society},
   volume = {111},
   number = {15},
   pages = {5649-5658},
   note = {doi: 10.1021/ja00197a022},
   ISSN = {0002-7863},
   DOI = {10.1021/ja00197a022},
   url = {https://doi.org/10.1021/ja00197a022},
   year = {1989},
   type = {Journal Article}
}

@article{berendsen1984bath,
  author    = {H. J. C. Berendsen and J. P. M. Postma and W. F. van Gunsteren and A. DiNola and J. R. Haak},
  title     = {Molecular dynamics with coupling to an external bath},
  journal   = {The Journal of Chemical Physics},
  volume    = {81},
  number    = {8},
  pages     = {3684-3690},
  year      = {1984},
  doi       = {10.1063/1.448118}
}

@article{g16,
author={M. J. Frisch and G. W. Trucks and H. B. Schlegel and G. E. Scuseria and M. A. Robb and J. R. Cheeseman and G. Scalmani and V. Barone and G. A. Petersson and H. Nakatsuji and X. Li and M. Caricato and A. V. Marenich and J. Bloino and B. G. Janesko and R. Gomperts and B. Mennucci and H. P. Hratchian and J. V. Ortiz and A. F. Izmaylov and J. L. Sonnenberg and D. Williams-Young and F. Ding and F. Lipparini and F. Egidi and J. Goings and B. Peng and A. Petrone and T. Henderson and D. Ranasinghe and V. G. Zakrzewski and J. Gao and N. Rega and G. Zheng and W. Liang and M. Hada and M. Ehara and K. Toyota and R. Fukuda and J. Hasegawa and M. Ishida and T. Nakajima and Y. Honda and O. Kitao and H. Nakai and T. Vreven and K. Throssell and Montgomery, {Jr.}, J. A. and J. E. Peralta and F. Ogliaro and M. J. Bearpark and J. J. Heyd and E. N. Brothers and K. N. Kudin and V. N. Staroverov and T. A. Keith and R. Kobayashi and J. Normand and K. Raghavachari and A. P. Rendell and J. C. Burant and S. S. Iyengar and J. Tomasi and M. Cossi and J. M. Millam and M. Klene and C. Adamo and R. Cammi and J. W. Ochterski and R. L. Martin and K. Morokuma and O. Farkas and J. B. Foresman and D. J. Fox},
title={Gaussian˜16 {R}evision {C}.01},
year={2016},
note={Gaussian Inc. Wallingford CT}
}

@article{onufriev2014opc,
author = {Izadi, Saeed and Anandakrishnan, Ramu and Onufriev, Alexey V.},
title = {Building Water Models: A Different Approach},
journal = {The Journal of Physical Chemistry Letters},
volume = {5},
number = {21},
pages = {3863-3871},
year = {2014},
doi = {10.1021/jz501780a},
note ={PMID: 25400877},
URL = {https://doi.org/10.1021/jz501780a},
eprint = {https://doi.org/10.1021/jz501780a}
}

@article{wang2020gaff2,
    author = {He, Xibing and Man, Viet H. and Yang, Wei and Lee, Tai-Sung and Wang, Junmei},
    title = {A fast and high-quality charge model for the next generation general AMBER force field},
    journal = {The Journal of Chemical Physics},
    volume = {153},
    number = {11},
    pages = {114502},
    year = {2020},
    month = {09},
    issn = {0021-9606},
    doi = {10.1063/5.0019056},
    url = {https://doi.org/10.1063/5.0019056},
    eprint = {https://pubs.aip.org/aip/jcp/article-pdf/doi/10.1063/5.0019056/15581389/114502\_1\_online.pdf},
}

@article{woods2000resp,
title = {Restrained electrostatic potential atomic partial charges for condensed-phase simulations of carbohydrates},
journal = {Journal of Molecular Structure: THEOCHEM},
volume = {527},
number = {1},
pages = {149-156},
year = {2000},
issn = {0166-1280},
doi = {https://doi.org/10.1016/S0166-1280(00)00487-5},
url = {https://www.sciencedirect.com/science/article/pii/S0166128000004875},
author = {R.J Woods and R Chappelle}
}

@article{darden1993pme,
    author = {Darden, Tom and York, Darrin and Pedersen, Lee},
    title = {Particle mesh Ewald: An N$\cdot$log(N) method for Ewald sums in large systems},
    journal = {The Journal of Chemical Physics},
    volume = {98},
    number = {12},
    pages = {10089-10092},
    year = {1993},
    month = {06},
    issn = {0021-9606},
    doi = {10.1063/1.464397},
    url = {https://doi.org/10.1063/1.464397},
    eprint = {https://pubs.aip.org/aip/jcp/article-pdf/98/12/10089/19327766/10089\_1\_online.pdf},
}

@misc{aleksandrowicz2019qiskit,
        title={ Qiskit: An Open-source Framework for Quantum Computing },
        author={ Gadi Aleksandrowicz and Thomas Alexander and Panagiotis Kl. Barkoutsos and Luciano Bello and Yael Ben-Haim and David Bucher and Francisco Jose Cabrera-Hernández and Jorge Carballo-Franquis and Adrian Chen and Chun-Fu Chen and Jerry M. Chow and Antonio D. Córcoles-Gonzales and Abigail J. Cross and Andrew W. Cross and Juan Cruz-Benito and Chris Culver and Salvador De La Puente González and Enrique De La Torre and Delton Ding and Eugene F. Dumitrescu and Ivan Duran and Pieter T. Eendebak and Mark Everitt and Ismael Faro Sertage and Albert Frisch and Andreas Fuhrer and Jay M. Gambetta and Borja Godoy Gago and Juan Gomez-Mosquera and Donny Greenberg and Ikko Hamamura and Vojtech Havlicek and Joe Hellmers and Łukasz Herok and Hiroshi Horii and Shaohan Hu and Takashi Imamichi and Toshinari Itoko and Ali Javadi-Abhari and Naoki Kanazawa and Anton Karazeev and Kevin Krsulich and Peng Liu and Yang Luh and Yunho Maeng and Manoel Marques and Francisco Martín-Fernández and Douglas McClure and David McKay and Srujan Meesala and Antonio Mezzacapo and Nikolaj Moll and Diego Moreda Rodríguez and Giacomo Nannicini and P. D. Nation and Pauline J. Ollitrault and Lee James O'Riordan and Hanhee Paik and Jesús Pérez and Anna Phan and Marco Pistoia and Viktor Prutyanov and Max Reuter and Julia E. Rice and Abdón Rodríguez Davila and Raymond Harry Rudy and Mingi Ryu and Ninad Sathaye and Chris Schnabel and Eddie Schoute and Kanav Setia and Yunong Shi and Adenilton Silva and Yukio Siraichi and Seyon Sivarajah and John A. Smolin and Mathias Soeken and Hitomi Takahashi and Ivano Tavernelli and Charles Taylor and Pete Taylour and Kenso Trabing and Matthew Treinish and Wes Turner and Desiree Vogt-Lee and Christophe Vuillot and Jonathan A. Wildstrom and Jessica Wilson and Erick Winston and Christopher J. Wood and Stephen P. Wood and Stefan Wörner and Ismail Yunus Akhalwaya and Christa Zoufal },        
        year={ 2019 },     
        doi={ 10.5281/ZENODO.2562111 },  
      }

@article{robledo2024chemistry,
doi = {10.1126/sciadv.adu9991},
author = {Javier Robledo-Moreno  and Mario Motta  and Holger Haas  and Ali Javadi-Abhari  and Petar Jurcevic  and William Kirby  and Simon Martiel  and Kunal Sharma  and Sandeep Sharma  and Tomonori Shirakawa  and Iskandar Sitdikov  and Rong-Yang Sun  and Kevin J. Sung  and Maika Takita  and Minh C. Tran  and Seiji Yunoki  and Antonio Mezzacapo },
title = {Chemistry beyond the scale of exact diagonalization on a quantum-centric supercomputer},
journal = {Science Advances},
volume = {11},
number = {25},
pages = {eadu9991},
year = {2025},
doi = {10.1126/sciadv.adu9991},
URL = {https://www.science.org/doi/abs/10.1126/sciadv.adu9991},
eprint = {https://www.science.org/doi/pdf/10.1126/sciadv.adu9991}
}

@article{holmes2016heat,
  author = {Holmes, A. A. and Tubman, N. M. and Umrigar, C.},
  title = {Heat-bath configuration interaction: An efficient selected configuration interaction algorithm inspired by heat-bath sampling},
  journal = {Journal of Chemical Theory and Computation},
  year = {2016},
  volume = {12},
  number = {8},
  pages = {3674-3680},
  doi = {10.1021/acs.jctc.6b00407}
}

@article{sharma2017semistochastic,
  title={Semistochastic heat-bath configuration interaction method: Selected configuration interaction with semistochastic perturbation theory},
  author={Sharma, S. and Holmes, A. A. and Jeanmairet, G. and Alavi, A. and Umrigar, C. J.},
  journal={Journal of Chemical Theory and Computation},
  year={2017},
  volume={13},
  number={4},
  pages={1595--1604},
  doi={10.1021/acs.jctc.6b01028}
}

@ARTICLE{shirakawalucj2025,
       author = {{Shirakawa}, Tomonori and {Robledo-Moreno}, Javier and {Itoko}, Toshinari and {Tripathi}, Vinay and {Ueda}, Kento and {Kawashima}, Yukio and {Broers}, Lukas and {Kirby}, William and {Pathak}, Himadri and {Paik}, Hanhee and {Tsuji}, Miwako and {Kodama}, Yuetsu and {Sato}, Mitsuhisa and {Evangelinos}, Constantinos and {Seelam}, Seetharami and {Walkup}, Robert and {Yunoki}, Seiji and {Motta}, Mario and {Jurcevic}, Petar and {Horii}, Hiroshi and {Mezzacapo}, Antonio},
        title = "{Closed-loop calculations of electronic structure on a quantum processor and a classical supercomputer at full scale}",
      journal = {arXiv e-prints},
         year = 2025,
        month = oct,
          eid = {arXiv:2511.00224},
        pages = {arXiv:2511.00224},
          doi = {10.48550/arXiv.2511.00224},
archivePrefix = {arXiv},
       eprint = {2511.00224},
 primaryClass = {quant-ph},
       adsurl = {https://ui.adsabs.harvard.edu/abs/2025arXiv251100224S}
}

@ARTICLE{exthci2025,
       author = {{Barroca}, Marco Antonio and {Gujarati}, Tanvi and {Sharma}, Vidushi and {Neumann Barros Ferreira}, Rodrigo and {Na}, Young-Hye and {Giammona}, Maxwell and {Mezzacapo}, Antonio and {Wunsch}, Benjamin and {Steiner}, Mathias},
        title = "{Scaling active spaces in simulations of surface reactions through sample-based quantum diagonalization}",
      journal = {arXiv e-prints},
         year = 2025,
        month = mar,
          eid = {arXiv:2503.10923},
        pages = {arXiv:2503.10923},
          doi = {10.48550/arXiv.2503.10923},
archivePrefix = {arXiv},
       eprint = {2503.10923},
 primaryClass = {quant-ph},
       adsurl = {https://ui.adsabs.harvard.edu/abs/2025arXiv250310923B}
}

@article{Sayfutyarova2017,
author = {Sayfutyarova, Elvira R. and Sun, Qiming and Chan, Garnet Kin-Lic and Knizia, Gerald},
title = {Automated Construction of Molecular Active Spaces from Atomic Valence Orbitals},
journal = {Journal of Chemical Theory and Computation},
volume = {13},
number = {9},
pages = {4063-4078},
year = {2017},
doi = {10.1021/acs.jctc.7b00128}
}

@article{sun2020recent,
  title={Recent developments in the PySCF program package},
  author={Sun, Q. and Zhang, X. and Banerjee, S. and Bao, P. and Barbry, M. and Blunt, N. S. and Bogdanov, N. A. and Booth, G. H. and Chen, J. and Cui, Z.-H.},
  journal={The Journal of Chemical Physics},
  year={2020},
  volume={153},
  number={2},
  pages={024109},
  doi={10.1063/5.0006074}
}

@article{sun2018pyscf,
  title={PySCF: the Python-based simulations of chemistry framework},
  author={Sun, Q. and Berkelbach, T. C. and Blunt, N. S. and Booth, G. H. and Guo, S. and Li, Z. and Liu, J. and McClain, J. D. and Sayfutyarova, E. R. and Sharma, S.},
  journal={Wiley Interdisciplinary Reviews: Computational Molecular Science},
  year={2018},
  volume={8},
  number={1},
  pages={e1340},
  doi={10.1002/wcms.1340}
}

@article{javadi2024quantum,
  title={Quantum computing with Qiskit},
  author={Javadi-Abhari, A. and Treinish, M. and Krsulich, K. and Wood, C. J. and Lishman, J. and Gacon, J. and Martiel, S. and Nation, P. D. and Bishop, L. S. and Cross, A. W.},
  year={2024},
  journal={arXiv preprint arXiv:2405.08810},
  note = {Available at \url{https://arxiv.org/abs/2405.08810}}
}

@article{motta2023bridging,
  title={Bridging physical intuition and hardware efficiency for correlated electronic states: the local unitary cluster Jastrow ansatz for electronic structure},
  author={Motta, M. and Sung, K. J. and Whaley, K. B. and Head-Gordon, M. and Shee, J.},
  journal={Chemical Science},
  year={2023},
  volume={14},
  number={40},
  pages={11213--11227},
  doi={10.1039/D3SC02609B}
}

@misc{ffsim2024,
  title={ffsim: Faster simulations of fermionic quantum circuits},
  author={{ffsims developers}},
  howpublished={\url{https://github.com/qiskit-community/ffsim}},
  year={2024},
  note={Accessed: 2024-09-01}
}

@article{7,
   author = {Pace, C. N. and Fu, H. and Fryar, K. L. and Landua, J. and Trevino, S. R. and Shirley, B. A. and Hendricks, M. M. and Iimura, S. and Gajiwala, K. and Scholtz, J. M. and Grimsley, G. R.},
   title = {Contribution of hydrophobic interactions to protein stability},
   journal = {J Mol Biol},
   volume = {408},
   number = {3},
   pages = {514-28},
   note = {},
   ISSN = {0022-2836 (Print)
0022-2836},
   DOI = {10.1016/j.jmb.2011.02.053},
   year = {2011},
   type = {Journal Article}
}

@article{10,
   author = {Hanshaw, R. G. and Stahelin, R. V. and Smith, B. D.},
   title = {Noncovalent keystone interactions controlling biomembrane structure},
   journal = {Chemistry},
   volume = {14},
   number = {6},
   pages = {1690-7},
   note = {},
   ISSN = {0947-6539 (Print)
0947-6539},
   DOI = {10.1002/chem.200701589},
   year = {2008},
   type = {Journal Article}
}

@article{14,
   author = {Yang, Y. and Zhang, H. and Wanyan, Y. and Liu, K. and Lv, T. and Li, M. and Chen, Y.},
   title = {Effect of Hydrophobicity on the Anticancer Activity of Fatty-Acyl-Conjugated CM4 in Breast Cancer Cells},
   journal = {ACS Omega},
   volume = {5},
   number = {34},
   pages = {21513-21523},
   note = {},
   ISSN = {2470-1343},
   DOI = {10.1021/acsomega.0c02093},
   year = {2020},
   type = {Journal Article}
}

@article{16,
   author = {Zwanzig, R. and Szabo, A. and Bagchi, B.},
   title = {Levinthal's paradox},
   journal = {Proc Natl Acad Sci USA},
   volume = {89},
   number = {1},
   pages = {20-2},
   note = {},
   ISSN = {0027-8424 (Print)
0027-8424},
   DOI = {10.1073/pnas.89.1.20},
   year = {1992},
   type = {Journal Article}
}

@article{17,
title = {The Levinthal paradox: yesterday and today},
journal = {Folding and Design},
volume = {2},
pages = {S69-S75},
year = {1997},
issn = {1359-0278},
doi = {https://doi.org/10.1016/S1359-0278(97)00067-9},
url = {https://www.sciencedirect.com/science/article/pii/S1359027897000679},
author = {Martin Karplus}
}

@article{18,
title = {Protein folding: from the Levinthal paradox to structure prediction},
journal = {Journal of Molecular Biology},
volume = {293},
number = {2},
pages = {283-293},
year = {1999},
issn = {0022-2836},
doi = {https://doi.org/10.1006/jmbi.1999.3006},
url = {https://www.sciencedirect.com/science/article/pii/S0022283699930061},
author = {Barry Honig}
}

@Article{ani1,
author ="Smith, J. S. and Isayev, O. and Roitberg, A. E.",
title  ="ANI-1: an extensible neural network potential with DFT accuracy at force field computational cost",
journal  ="Chem. Sci.",
year  ="2017",
volume  ="8",
issue  ="4",
pages  ="3192-3203",
publisher  ="The Royal Society of Chemistry",
doi  ="10.1039/C6SC05720A",
url  ="http://dx.doi.org/10.1039/C6SC05720A"}

@article{ani2,
author = {Devereux, Christian and Smith, Justin S. and Huddleston, Kate K. and Barros, Kipton and Zubatyuk, Roman and Isayev, Olexandr and Roitberg, Adrian E.},
title = {Extending the Applicability of the ANI Deep Learning Molecular Potential to Sulfur and Halogens},
journal = {Journal of Chemical Theory and Computation},
volume = {16},
number = {7},
pages = {4192-4202},
year = {2020},
doi = {10.1021/acs.jctc.0c00121},
    note ={PMID: 32543858},
URL = {https://doi.org/10.1021/acs.jctc.0c00121},
eprint = {https://doi.org/10.1021/acs.jctc.0c00121}
}

@article{ani2024,
   author = {Rezaee, Mozafar and Ekrami, Saeid and Hashemianzadeh, Seyed Majid},
   title = {Comparing ANI-2x, ANI-1ccx neural networks, force field, and DFT methods for predicting conformational potential energy of organic molecules},
   journal = {Scientific Reports},
   volume = {14},
   number = {1},
   pages = {11791},
   ISSN = {2045-2322},
   DOI = {10.1038/s41598-024-62242-5},
   url = {https://doi.org/10.1038/s41598-024-62242-5},
   year = {2024},
   type = {Journal Article}
}

@article{Gao1992,
doi = {10.1021/j100181a009},
author = {Gao, Jiali},
title = {Absolute free energy of solvation from Monte Carlo simulations using combined quantum and molecular mechanical potentials},
journal = {The Journal of Physical Chemistry},
volume = {96},
number = {2},
pages = {537-540},
year = {1992},
doi = {10.1021/j100181a009},
URL = {https://doi.org/10.1021/j100181a009},
eprint = {https://doi.org/10.1021/j100181a009}}

@article{gao1992priori,
  title={A priori evaluation of aqueous polarization effects through Monte Carlo QM-MM simulations},
  author={Gao, Jiali and Xia, Xinfu},
  journal={Science},
  volume={258},
  number={5082},
  pages={631--635},
  year={1992},
  month={oct},
  publisher={American Association for the Advancement of Science},
  doi={10.1126/science.1411573},
  pmid={1411573}}

@article{Warshel1992,
       author = {{Luzhkov}, V. and {Warshel}, A.},
        title = "{Microscopic models for quantum mechanical calculations of chemical processes in solutions: LD/AMPAC and SCAAS/AMPAC calculations of solvation energies}",
      journal = {Journal of Computational Chemistry},
         year = 1992,
        month = mar,
       volume = {13},
       number = {2},
        pages = {199-213},
          doi = {10.1002/jcc.540130212},
       adsurl = {https://ui.adsabs.harvard.edu/abs/1992JCoCh..13..199L}}

@article{hudson2018force,
  title={Force matching as a stepping stone to QM/MM CB[8] host/guest binding free energies: a SAMPL6 cautionary tale},
  author={Hudson, Phillip S. and Han, Kyungreem and Woodcock, H. Lee and Brooks, Bernard R.},
  journal={Journal of Computer-Aided Molecular Design},
  volume={32},
  number={10},
  pages={983--999},
  year={2018},
  month={oct},
  publisher={Springer},
  doi={10.1007/s10822-018-0165-3},
  pmid={30276502},
  pmcid={PMC6867086}}

@article{konig2014,
doi = {10.1021/ct401118k},
author = {König, Gerhard and Hudson, Phillip S. and Boresch, Stefan and Woodcock, H. Lee},
title = {Multiscale Free Energy Simulations: An Efficient Method for Connecting Classical MD Simulations to QM or QM/MM Free Energies Using Non-Boltzmann Bennett Reweighting Schemes},
journal = {Journal of Chemical Theory and Computation},
volume = {10},
number = {4},
pages = {1406-1419},
year = {2014},
doi = {10.1021/ct401118k},
note ={PMID: 24803863},
URL = {https://doi.org/10.1021/ct401118k},
eprint = {https://doi.org/10.1021/ct401118k}}

@article{konig2014multiscale,
doi = {10.1021/ct401118k},
author = {König, Gerhard and Hudson, Phillip S. and Boresch, Stefan and Woodcock, H. Lee},
title = {Multiscale Free Energy Simulations: An Efficient Method for Connecting Classical MD Simulations to QM or QM/MM Free Energies Using Non-Boltzmann Bennett Reweighting Schemes},
journal = {Journal of Chemical Theory and Computation},
volume = {10},
number = {4},
pages = {1406-1419},
year = {2014},
doi = {10.1021/ct401118k},
note ={PMID: 24803863},
URL = {https://doi.org/10.1021/ct401118k},
eprint = { https://doi.org/10.1021/ct401118k}}

@article{konig2015,
  title={Correcting for the free energy costs of bond or angle constraints in molecular dynamics simulations},
  author={König, Gerhard and Brooks, Bernard R.},
  journal={Biochimica et Biophysica Acta (BBA) - General Subjects},
  volume={1850},
  number={5},
  pages={932--943},
  year={2015},
  month={may},
  publisher={Elsevier},
  doi={10.1016/j.bbagen.2014.09.001},
  pmid={25218695},
  pmcid={PMC4339525}}

@article{konig2018,
  title={On the convergence of multi-scale free energy simulations},
  author={König, Gerhard and Brooks, Bernard R. and Thiel, Walter and York, Darrin M.},
  journal={Molecular Simulation},
  volume={44},
  number={13-14},
  pages={1062--1081},
  year={2018},
  publisher={Taylor \& Francis},
  doi={10.1080/08927022.2018.1475741},
  pmid={30581251},
  pmcid={PMC6298030}
}

@article{hudson2018accelerating,
  title={Accelerating QM/MM Free Energy Computations via Intramolecular Force Matching},
  author={Hudson, Phillip S. and Boresch, Stefan and Rogers, David M. and Woodcock, H. Lee},
  journal={Journal of Chemical Theory and Computation},
  volume={14},
  number={12},
  pages={6327--6335},
  year={2018},
  month={dec},
  publisher={American Chemical Society},
  doi={10.1021/acs.jctc.8b00517},
  pmid={30300543},
  pmcid={PMC6314469}}

@article{hudson2015,
doi = {10.1021/acs.jpclett.5b02164},
author = {Hudson, Phillip S. and Woodcock, H. Lee and Boresch, Stefan},
title = {Use of Nonequilibrium Work Methods to Compute Free Energy Differences Between Molecular Mechanical and Quantum Mechanical Representations of Molecular Systems},
journal = {The Journal of Physical Chemistry Letters},
volume = {6},
number = {23},
pages = {4850-4856},
year = {2015},
doi = {10.1021/acs.jpclett.5b02164},
note ={PMID: 26539729},
URL = {https://doi.org/10.1021/acs.jpclett.5b02164},
eprint = {https://doi.org/10.1021/acs.jpclett.5b02164}}

@article{konig2018comparison,
  title={A Comparison of QM/MM Simulations with and without the Drude Oscillator Model Based on Hydration Free Energies of Simple Solutes},
  author={König, Gerhard and Pickard, Frank C. and Huang, Jing and Thiel, Walter and MacKerell, Alexander D. and Brooks, Bernard R. and York, Darrin M.},
  journal={Molecules},
  volume={23},
  number={10},
  pages={2695},
  year={2018},
  month={oct},
  publisher={MDPI},
  doi={10.3390/molecules23102695},
  pmid={30347691},
  pmcid={PMC6222909}}

@article{Kearns2017,
author = {Kearns, Fiona L. and Hudson, Phillip S. and Woodcock, Henry L. and Boresch, Stefan},
title = {Computing converged free energy differences between levels of theory via nonequilibrium work methods: Challenges and opportunities},
journal = {Journal of Computational Chemistry},
volume = {38},
number = {16},
pages = {1376-1388},
doi = {https://doi.org/10.1002/jcc.24706},
url = {https://onlinelibrary.wiley.com/doi/abs/10.1002/jcc.24706},
year = {2017}}

@article{Gao2024,
author = {Gao, Hong and Imamura, Satoshi and Kasagi, Akihiko and Yoshida, Eiji},
title = {Distributed Implementation of Full Configuration Interaction for One Trillion Determinants},
journal = {Journal of Chemical Theory and Computation},
volume = {20},
number = {3},
pages = {1185-1192},
year = {2024},
doi = {10.1021/acs.jctc.3c01190},
    note ={PMID: 38314701},

URL = { 
    
        https://doi.org/10.1021/acs.jctc.3c01190
    
    

},
eprint = { 
    
        https://doi.org/10.1021/acs.jctc.3c01190
    
    

}

}

@article{Vogiatzis2017,
    author = {Vogiatzis, Konstantinos D. and Ma, Dongxia and Olsen, Jeppe and Gagliardi, Laura and de Jong, Wibe A.},
    title = "{Pushing configuration-interaction to the limit: Towards massively parallel MCSCF calculations}",
    journal = {The Journal of Chemical Physics},
    volume = {147},
    number = {18},
    pages = {184111},
    year = {2017},
    month = {11},
    issn = {0021-9606},
    doi = {10.1063/1.4989858},
    url = {https://doi.org/10.1063/1.4989858},
    eprint = {https://pubs.aip.org/aip/jcp/article-pdf/doi/10.1063/1.4989858/13501842/184111\_1\_online.pdf},
}

@article{shajan2024towards,
  title={Towards quantum-centric simulations of extended molecules: sample-based quantum diagonalization enhanced with density matrix embedding theory},
  author={Shajan, Akhil and Kaliakin, Danil and Mitra, Abhishek and Moreno, Javier Robledo and Li, Zhen and Motta, Mario and Johnson, Caleb and Saki, Abdullah Ash and Das, Susanta and Sitdikov, Iskandar and others},
  journal={arXiv preprint arXiv:2411.09861},
  year={2024}
}

@article{kaliakin2024accurate,
  title={Accurate quantum-centric simulations of supramolecular interactions},
  author={Kaliakin, D. and Shajan, A. and Moreno, J. R. and Li, Z. and Mitra, A. and Motta, M. and Johnson, C. and Saki, A. A. and Das, S. and Sitdikov, I.},
  journal={arXiv preprint arXiv:2410.09209},
  year={2024},
  url={https://arxiv.org/abs/2410.09209}
}

\setcounter{figure}{0} 
\renewcommand{\thefigure}{S\arabic{figure}}

\newpage
\begin{center}
\section*{Supplementary Information: Protein-Ligand Free Energy Perturbation on Quantum Hardware}
\end{center}

\subsection{Multistate Bennett Acceptance Ratio (MBAR) analysis}
The MBAR calculations \cite{mbar} are often derived from 
\begin{equation}\label{eq4}
    \Delta A_{ij} = \frac{\ln\frac{Q_j}{Q_i}}{-\beta}
    \;.
\end{equation}

In Eq.~\eqref{eq4}, $\Delta A_{ij}$ denotes the relative Helmholtz free energy difference between two designated states, $\lambda_i$, $\lambda_j$, for the multistate system, where $Q_i$ and $Q_j$ denote the partition functions of the two states, as expressed in the Poisson-Boltzmann distribution. Here, $\beta$ is defined as $1/(k_BT)$ , where $k_B$ is the Poisson-Boltzmann constant and $T$ is the absolute temperature. Re-writing Eq.~\eqref{eq4}, we obtain
\begin{equation}\label{eq5}
    Q_i\langle\alpha_{ij}\times e^{-\beta U_j}\rangle_i = Q_j\langle\alpha_{ij}\times e^{-\beta U_i}\rangle_j 
\end{equation}

In Eq.~\eqref{eq5}, $\alpha_{ij}$ is the probability that a sample drawn from state $\lambda_j$ appears to follow the distribution of partition functions from state $\lambda_i$. A larger $\alpha_{ij}$ value usually means a better overlap between states $\lambda_i$ and $\lambda_j$. When sampling is sufficient, according to the definition of $\langle\alpha_{ij}\times e^{-\beta U_j}\rangle_i$, $\alpha_{ij}$ can be further expressed as below in an $i$-independent way

\begin{equation}\label{eq6}
    \alpha_{ij} = \frac{\frac{N_j}{\hat{c}_j}}{\sum\limits_{k=1}^K \frac{N_k  e^{\left(-\beta U_k \right)}}{\hat{c}_k}}
\end{equation}

In Eq.~\eqref{eq6}, $K$ is the total number of $\lambda_k$ states (in this work, we selected six, which are from $\lambda_1=0.0$, $\lambda_2=0.2$ ... to $\lambda_6=1.0$). $N_k$ denotes the total number of samples drawn at state $\lambda_k$, and $\hat{c_k}$ denotes the partition function of state 
$\lambda_k$ up to a constant (which is the arbitrarily defined acceptance ratio in MBAR).

Inserting Eq.~\eqref{eq6} back to Eq.~\eqref{eq5} and finally to Eq.~\eqref{eq4}, we have the self consistency expression of 

\begin{equation}\label{eq7}
     \hat{A_i} = \frac{\ln \sum\limits_{j=1}^K \sum\limits_{i=1}^{N_j} \frac{e^{-\beta U_i}}{\sum_{k=1}^K N_k e^{\beta\hat{A_k} - \beta U_k}}}{-\beta}
\end{equation}

In Eq.~\eqref{eq7}, once we have an initial guess of $\{\hat{A_k}, k\in \{1,2,3,...,6\}\})$ since we have six $\lambda$ values, we will then be able to iteratively update each $\hat{A_k}$ using Eq.~\eqref{eq4} until the optimized Helmholtz free energy is obtained in the NVT ensemble, then converted to the Gibbs free energy in the NPT ensemble. The whole calculation and analysis is automated via the AmberTools25 package~\cite{amber25}. The final corrected book-ending energy $\Delta \Delta G_{book-ending}^{QM/MM}$ is then added on top of the MM HFE as Figure~\ref{fig1} shows. 

\subsection{QUICK interface with PySCF and Qiskit Addon: SQD-extSQD}

To enable coupling between MD simulations performed on classical hardware and quantum circuit-based computations, we developed a custom interface that connects AMBER/QUICK with Qiskit \cite{aleksandrowicz2019qiskit, javadi2024quantum}. In addition, we integrated simulations with PySCF 2.8.0 \cite{sun2018pyscf,sun2020recent} into the AMBER/QUICK workflow to enable HCI calculations (Figure S1). The standard API between AMBER \cite{amber25} and QUICK \cite{manathunga2023quantum,cruzeiro2021open} was extended to include both two new modules: 
A) the HCI solver, implemented through the interface with PySCF, and qiskit-addon-dice-solver, and 
B) the SQD-extSQD solver, which utilizes the quantum hardware to generate the initial electron configurations with LUCJ Ansatz and subsequent post-processing through qiskit-addon-sqd and PyCI for augmentation of the Hilbert space. 

The AMBER Fortran module, specifically quick\_api\_module.F90, was modified to interface the classical engine (AMBER/QUICK) with the PySCF code, managing data exchange between the classical simulation and the quantum backends. A dedicated keyword CI\_stride was specified within AMBER to define the frequency (in MD steps) at which the HCI or SQD-extSQD solver is triggered. QUICK's API mode supports user-defined QM methods, basis sets, and Molden export via the AMBER MD input file. 

Shared functionality is implemented in the quick\_parsing\_utilities.py script. This script reads the quick.out file, extracts the coordinates and any associated MM point charges, forwards them to the external solver, and finally integrates the computed energy, nuclear gradient, and correlation energy into a new output file, QUICK\_job\_PySCF\_modified.out. When external molecules surround the QM region, their coordinates and charges are passed to the molecular system definition within PySCF, and the system is initialized using the PySCF QM/MM module, enabling electrostatic embedding of the classical environment. Once the HF object is constructed, it is populated with orbital coefficients, energies, and occupations extracted from the Molden file. In the HCI solver, an HCI calculation is performed using  CI\_solver\_PySCF.py, and the result is integrated into AMBER through QUICK\_job\_PySCF\_modified.out, influencing the subsequent MD propagation. For the SQD solver, the quantum-centric simulation, two Python scripts are provided. LUCJ-run.py prepares the quantum input by saving the one-electron and two-electron integrals of the orbitals within the active space. These integrals are written in a fcidump file together with the nuclear repulsion and the electron number. In the following step, the fcidump is read back to run a Coupled Cluster Singles and Doubles (CCSD) calculation, from which the single t\textsubscript{1} and double t\textsubscript{2} excitation amplitudes are extracted. 

\begin{figure}
    \centering
    \includegraphics[width=1\linewidth]{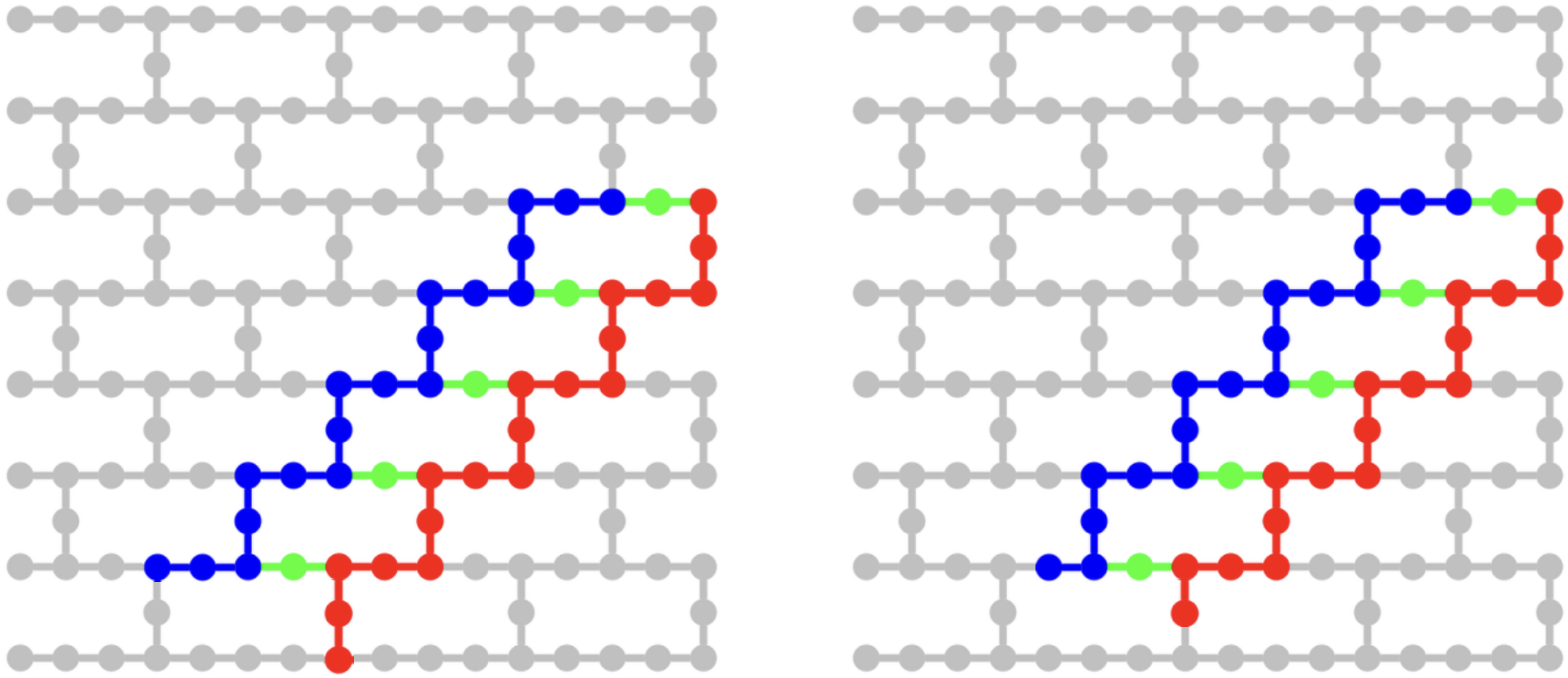}
    \caption{
    Qubits layout of the LUCJ circuit, both thermolysin inhibitors in the present study (0PJ: 26e, 19o; 0PI: 26e, 18o), after Cartesian-based AVAS orbital selection. Qubits corresponding to the occupation numbers of $\alpha$ and $\beta$ spin-orbitals are represented in red and blue, respectively. Auxiliary qubits employed to mediate the density-density interactions between $\alpha$ and $\beta$ spin-orbitals are depicted in green. Timing-wise, each LUCJ circuit takes 1 minute to execute, yielding a total of 2400 minutes of hardware time, given 50 CI\_stride steps for each of the 4 ligand/protein combinations, with 6 lambda windows and 2 duplicates.}
    \label{figS1}
\end{figure}

\begin{figure}
    \centering
    \includegraphics[width=1\linewidth]{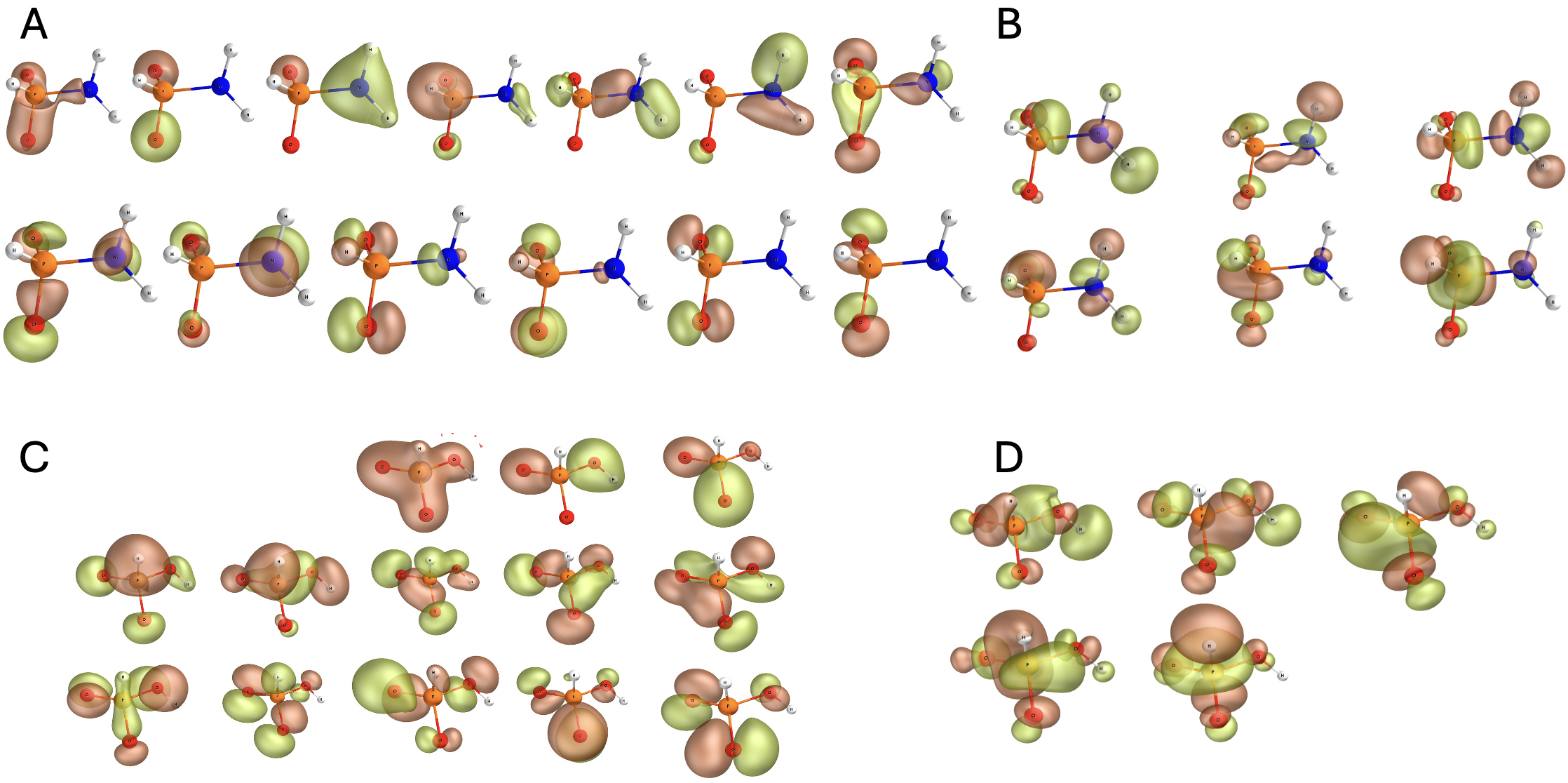}
    \caption{
    Active spaces of the two inhibitors of thermolysin. (A) Occupied and (B) Virtual orbitals of the 0PJ inhibitor. (C) Occupied and (D) Virtual orbitals of the 0PI inhibitor, which has one less hydrogen bond compared to 0PJ while inhibiting thermolysin.}
    \label{figS2}
\end{figure}

We generate the LUCJ circuits using the ffsim library (version 0.0.49) \cite{ffsim2024} interfaced with Qiskit 1.3.3 \cite{aleksandrowicz2019qiskit,javadi2024quantum}. The t\textsubscript{2} amplitudes are used to parametrize the LUCJ ansatz, constructed via the \texttt{UCJOpSpinBalanced} class. The circuits include 2 repetitions (\texttt{n\_reps=2}) and entangling connections defined by nearest-neighbor and on-site interaction patterns in the active space. In the present work, we used \texttt{ibm\_marrakesh} quantum processor. Circuit transpilation is performed with optimization level 3 and includes a custom pre-initialization step provided by ffsim. Quantum error mitigation is applied through gate twirling (while measurement twirling is disabled), as enabled by the \texttt{SamplerV2} primitive in Qiskit's runtime library (version 0.36.1). The final circuit layout is reported in Figure~\ref {figS1} and the related active space is in Figure~\ref {figS2}.

The post-processing stage is managed by the scripts run\_SQD.py and ext\_SQD\_run.py, which utilize the qiskit-addon-sqd 0.11.0 library and read input data from count\_dict.txt. The output consists of a collection of quantum states (bitstrings) generated on the quantum hardware. Because of the noise inherent to current quantum devices, these bitstrings include both physical and non-physical configurations. To refine the dataset, the scripts first filter bitstrings with the correct particle number, corresponding to valid initial orbital occupations.

Subsequently, the self-consistent configuration recovery (S-CORE) procedure ~\cite{robledo2024chemistry} is applied. This algorithm iteratively corrects the spin-up and spin-down electron occupations in noise-contaminated bitstrings by comparing each bit’s current value to the average spin-orbital occupancy, thereby recovering physically meaningful configurations.

From this cleaned ensemble, batches of configurations are created to define subspaces within the full Hilbert space. For each subspace, the Hamiltonian is projected and diagonalized to obtain the ground state. The resulting states are analyzed to compute average orbital occupations, which guide the probabilistic generation of a new, refined ensemble of bitstrings. This iterative cycle—subspace construction, Hamiltonian projection, diagonalization, occupation averaging, and configuration recovery—continues until a convergence criterion is satisfied, progressively reducing noise and improving the accuracy of the quantum simulation results.

In this study, we employed 10 batches of 1000 samples each, parallelized across 10 CPUs using Slurm, and performed 2 S-CORE iterations and 1 extSQD iteration. To further extend the functionality, we introduced a monkey patch ~\cite{Hunt2023} to incorporate custom functions into the PySCF code. Specifically, we modified the CI-wrapper to integrate the kernel\_fixed\_space routine, allowing access to the gradient module of PySCF ~\cite{sun2018pyscf, sun2020recent}. This module, which supports analytical gradients for Selected CI (SCI) calculations ~\cite{Coe2023, Libcint15}, was extended to SQD-extSQD calculations in the present work.

The gradient was computed and incorporated into the molecular dynamics (MD) simulation only in the last extSQD step, corresponding to the optimal subspace obtained after configuration recovery. All simulations in this study employed all orbitals within the 6-31+G* basis set, where Cartesian-based AVAS selects MOs associated with 2s and 2pxyz of oxygens and nitrogens, plus 3s and 3pxyz orbitals of phosphorus in the QM region. 

\subsection{Modification on the AVAS code}

The modification addresses a PySCF bug triggered when Cartesian, rather than spherical, functions are used.

In detail, the original AVAS implementation finds the designated labels of the orbitals (\texttt{avas\_obj.aolabels} in \texttt{local\_avas.py}) in the list of labels of orbitals or a \texttt{pmol} molecular object (\texttt{pmol.ao\_labels()} in \texttt{local\_avas.py}). The result was assigned to a variable named \texttt{baslst}, which would be empty unless spherical functions were used. This is due to an inconsistency between how the labels are written in \texttt{avas\_obj.aolabels} and \texttt{pmol.ao\_labels()} when Cartesian functions are used, so they cannot be detected. 

Therefore, the modified texttt{local\_avas.py} ensures that \texttt{pmol} also uses Cartesian, although it uses spherical by default. Then, the modified \texttt{local\_avas.py} code converts the orbital labels to a common format before PySCF searches for the \texttt{baslst} variable.

\subsection{Quantum-centric SQD-extSQD simulations}

At the heart of SQD-extSQD \cite{robledo2024chemistry, kaliakin2024accurate, yu2025quantum, Liepuoniute2024, Barison2025, barroca2025surface, yu2025quantum} lies the execution of a quantum circuit designed to sample a set of states $\chi = \{ \bts_1 \dots \bts_d \}$  that serve as the computational basis for the diagonalization of the molecular Hamiltonian. 

An initial wavefunction guess $ \left| \Phi_{qc} \right\rangle $, approximating the ground state,  is prepared from a truncated version of the LUCJ ansatz. Here, the mapping of fermions to qubits is performed with the standard Jordan-Wigner (JW) mapping, and the LUCJ ansatz \cite{motta2023bridging} has the following form:
\begin{align}
|\Phi_{\mathrm{qc}}\rangle = e^{-\hat{K}_2} e^{\hat{K}_1} e^{i\hat{J}_1} e^{-\hat{K}_1} |x_{\mathrm{RHF}}\rangle,
\end{align}
where the one-body operators are denoted by $ \hat{K}_1 $ and $ \hat{K}_2$ , while $ \hat{J}_1 $ represents the density-density operator, and $|x_{\mathrm{RHF}}\rangle$ is the restricted closed-shell RHF state.  The LUCJ ansatz was parametrized using amplitudes obtained from the gas-phase restricted closed-shell CCSD calculations performed within a selected active space, following the approach adopted in prior quantum-centric SQD studies \cite{robledo2024chemistry, kaliakin2024accurate,yu2025quantum, Liepuoniute2024, Barison2025, barroca2025surface,yu2025quantum}. We run the quantum circuit on a quantum computer and measure the state \( |\Psi\rangle \) in the computational basis. By repeating the measurement many times, we collect a set of bitstrings.
\[
\tilde{\chi} = \{ \mathbf{x} \mid \mathbf{x} \sim \tilde{P}_\Psi(\mathbf{x}) \},
\]

Where each bitstring \( \mathbf{x} \in \{0,1\}^M \) represents an electronic configuration (Slater determinant) sampled according to \( \tilde{P}_\Psi \), however, the results obtained from the execution of a quantum circuit are inherently affected by errors arising from noise present in current quantum devices. This noise can introduce broken particle-number and spin-z symmetries in the samples, spreading the distribution  \( {P}_\Psi \) over configurations that do not contribute to the low-energy states.  As a consequence, only a fraction of the $\widetilde{\chi}$ contribute to the ground state. To mitigate this, we applied a configuration recovery protocol that iteratively recovers configurations with the correct particle number. For each configuration \( \mathbf{x} \in \tilde{\chi} \) such that \( N_{\mathbf{x}} \neq N \), a number \( |N_{\mathbf{x}} - N| \) of spin-orbitals are flipped. Bit-flip probabilities are weighted according to a monotonically increasing function of the difference $|x_{p\sigma} - n_{p\sigma}|$ where  $x_{p\sigma}$  is the bit value and  $n_{p\sigma}$  is the average occupation of spin-orbital $p\sigma$  from the previous recovery round. The initial guess of the occupancies n used in the first recovery iteration is computed from the raw quantum samples  $\widetilde{\chi}$  \cite{robledo2024chemistry}. From this process, we obtain the refined configurations $\chi_R$ from which we build K batches of d configurations, $S(1), \ldots, S(K)$, according to a distribution proportional to the empirical frequencies of each x in $\chi_R$.  Each batch spans a many-body subspace over which the Hamiltonian is projected and diagonalized:

\begin{equation}
\hat{H}_{S^{(b)}} = \hat{P}_{S^{(b)}} \hat{H} \hat{P}_{S^{(b)}},
\end{equation}

Where the projector $\hat{P}_{S^{(b)}}$ is,
\begin{equation}
\hat{P}_{S^{(b)}} = \sum_{x \in S^{(b)}} |x\rangle \langle x|.
\end{equation}

The diagonalization of  $\hat{H}_{S^{(b)}}$ yields the subspace ground state $|\psi^{(b)}\rangle$ and corresponding energy $E^{(b)}$, computed using the iterative Davidson method. 

We use the lowest energy across batches, $min_{b} E^{(b)}$, as the best approximation to the ground-state energy. The wavefunctions $|\psi^{(b)}\rangle$ from all the batches are then used to get the new occupancies,
\begin{equation}
n_{p\sigma} = \frac{1}{K} \sum_{b=1}^{K} \langle \psi^{(b)} | \hat{n}_{p\sigma} | \psi^{(b)} \rangle
\end{equation}

These occupancies are fed back into the configuration recovery step. The self-consistent procedure is repeated until convergence is reached. 

For the extSQD part, once the SQD iteration is complete, we further extend the subspace by applying the excitation operators to the configurations in the original subspace.

\begin{equation}\label{eq12}
     |\tilde{\Phi}_\mu\rangle=\sum_{I,k} d_{I\mu}\,c_{k0}\,\hat E_I\,|y_k\rangle
\end{equation}

Here $c_{k0}$ and $|y_k\rangle$ stand for the ground state coefficient and wave function, which fulfill the properties below:

\begin{equation}\label{eq13}
|\tilde{\Phi}_0\rangle=\sum_{k=1}^{D} c_{k0}\,|y_k\rangle,
\qquad
H_{lk}=\langle y_l|\hat H|y_k\rangle,
\qquad
\sum_{k=1}^{D} H_{lk} c_{k0}=\tilde{\varepsilon}_0\, c_{l0}
\end{equation}

And the other part of Eq.~\eqref{eq12} is associated with the single and double excitation. Given the simulation speed, we used only a single excitation in our work. 
\[
|\tilde{\Phi}_\mu\rangle=\sum_{I} d_{I\mu}\,\hat E_I\,|\tilde{\Phi}_0\rangle,
\quad
\hat E_I \in \Big\{\mathbf{1},\ \hat a^\dagger_{a\sigma}\hat a_{i\sigma},\
\hat a^\dagger_{a\sigma}\hat a^\dagger_{b\tau}\hat a_{j\tau}\hat a_{i\sigma}\Big\} \tag{14}
\]

Hence, the excitation of Eq.~\eqref{eq12} can be re-written as:
\begin{equation}\label{eq15}
\hat E_I|y_k\rangle=\gamma_{Ik}\,|z_{Ik}\rangle, \qquad \gamma_{Ik}\in\{-1,0,1\} 
\end{equation}

Plugging Eq.~\eqref{eq15} back to Eq.~\eqref{eq12} will yield

\[
|\tilde{\Phi}_\mu\rangle=\sum_{I,k} d_{I\mu}\,\gamma_{Ik}\,c_{k0}\,|z_{Ik}\rangle \tag{16}
\]

The boundary on the size of the extended space, D prime, can be expressed as
\[
D^{prime}=D_E=|S_E| \;\le\; D \prod_{\sigma\in\{\alpha,\beta\}} N_\sigma\,(M-N_\sigma) \tag{17}
\]
Where D is the original SQD subspace size, $M$ is the number of orbitals, and $N_\sigma$ is the electron's population in the spin orbital.

Converting the wave function to bitstring formats obtained from the variational ansatz, we finally have
\[
|\tilde{\Phi}_\mu\rangle=\sum_{z\in S_E} f_{z\mu}\,|z\rangle,
\qquad
\sum_{z'} \langle z|\hat H|z'\rangle\, f_{z'\mu}=\tilde{\varepsilon}_\mu\, f_{z\mu} \tag{18}
\]

Which can reduce the $\tilde{\varepsilon}_\mu$ value compared to the ground state eigenvalue $\tilde{\varepsilon}_0$ as mentioned in Eq.~\eqref{eq13}.

\end{document}